\documentclass[11pt%
amsmath,amssymb,
aps,
pra,
longbibliography
]{revtex4-1}
\usepackage[dvipsnames]{xcolor}
\usepackage{graphicx}								    
\usepackage{epstopdf,epsfig,mathtools}
\usepackage{dcolumn}								    
\usepackage{bm}									        
\usepackage{todonotes}
\newcommand{\mL}{\mathcal{L}}
\newcommand{\mJ}{\mathcal{J}}
\newcommand{\mN}{\mathcal{N}}
\newcommand{\oU}{\overline{U}}
\newcommand{\Temp}{\Theta}

\newcommand{\lap}{\nabla^2}
\newcommand{\lapd}{\nabla^4}
\newcommand{\bey}{\mathbf{e_y}}
\newcommand{\hu}{\hat{u}}
\newcommand{\ULC}{U_{\rm lc}}
\newcommand{\uR}{u_{\rm rzif}}

\newcommand{\uS}{u_{\rm scm}}
\newcommand{\US}{U_{\rm scm}}
\newcommand{\Ub}{U_{\rm b}}
\newcommand{\ub}{u_{\rm b}}

\newcommand{\sR}{\sigma_{\rm rzif}}
\newcommand{\sS}{\sigma_{\rm scm}}
\newcommand{\sL}{\sigma_{\rm b}}
\newcommand{\TLC}{T_{\rm lc}}
\newcommand{\oLC}{\omega_{\rm lc}}
\newcommand{\oR}{\omega_{\rm rzif}}
\newcommand{\oS}{\omega_{\rm scm}}
\newcommand{\oL}{\omega_{\rm b}}
\newcommand{\pd}{\partial}
\newcommand{\dt}{\Delta t}

\begin{document}

 \title{Frequency prediction from exact or self-consistent meanflows}\

\author{Yacine Bengana}
\email{b.y.bengana@gmail.com}
\affiliation{%
Department of Aeronautics, Imperial College London,
South Kensington, London SW7 2AZ, United Kingdom}

\author{Laurette S. Tuckerman}
\email{Laurette.Tuckerman@espci.fr}
\affiliation{%
Physique et M\'ecanique des Milieux H\'et\'erog\`enes, CNRS, ESPCI Paris, Universit\'e PSL, Sorbonne Universit\'e, Universit\'e de Paris, 75005 Paris, France}
\date{\today}											


\begin{abstract}
     
  A number of approximations have been proposed to estimate basic
  hydrodynamic quantities, in particular the frequency of a limit
  cycle. One of these, RZIF (for Real Zero Imaginary Frequency), calls
  for linearizing the governing equations about the mean flow and
  estimating the frequency as the imaginary part of the leading
  eigenvalue. A further reduction, the SCM (for Self-Consistent
  Model), approximates the mean flow as well, as resulting only from
  the nonlinear interaction of the leading eigenmode with itself.
  Both RZIF and SCM have proven dramatically successful for the
  archetypal case of the wake of a circular cylinder.

  Here, the SCM is applied to thermosolutal convection, for which a
  supercritical Hopf bifurcation gives rise to branches of standing
  waves and traveling waves. The SCM is solved by means of a full
  Newton method coupling the approximate mean flow and leading
  eigenmode.  Although the RZIF property is verified for the traveling
  waves, the SCM reproduces the nonlinear frequency only very near the
  onset of the bifurcation and for another isolated parameter value.
  Thus, the nonlinear interaction arising from the
  leading mode is insufficient to reproduce the nonlinear mean field
  and frequency.
\end{abstract}

\maketitle


\section{\label{sec:Intro}Introduction}

Periodic emission or translation of vortical structures is one
of the most important phenomena observed in hydrodynamic configurations.
The amplitude and frequency are two
essential characteristics of these time-periodic systems. 
These are usually obtained either by experiment or by solving the full
Navier-Stokes equations by direct numerical simulation.
The archetype of such configurations is the wake of a circular cylinder,
in which the visually appealing B\'enard-von-K\'arm\'an vortex ``street''
\cite{benard1908a,karman1911}
appears above a Reynolds-number threshold \cite{jackson1987finite,
provensal1987benard} of 46.

When periodic oscillations such as these originate from a supercritical Hopf
bifurcation, linear stability analysis about the
equilibrium solution at the threshold yields a leading eigenvalue
whose real part is zero and whose imaginary part is the oscillation
frequency.
Away from the threshold, this is no longer the case.  However, for the
cylinder wake, linearization about the {\it time-averaged field} has
been shown to yield the nonlinear frequency
\cite{pier2002frequency,barkley2006linear,sipp2007global,mittal2008global}
as the imaginary part of the leading eigenvalue.  Moreover Barkley
\cite{barkley2006linear} noted that the real part of this eigenvalue
is nearly zero, meaning that the mean flow can be considered to be
marginally stable, as had been suggested by Malkus
\cite{malkus1956outline} in the context of turbulent shear flow.  This
property, given the name RZIF for Real Zero Imaginary Frequency by
Turton, Tuckerman \& Barkley~\cite{turton2015}, has since been
demonstrated to hold for several other flow configurations, namely
traveling waves in thermosolutal convection \cite{turton2015}, spirals
and ribbons in counter-rotating Taylor-Couette flow
\cite{bengana2019spirals}, and (approximately) for the flow in a
two-dimensional shear-driven cavity
\cite{sipp2007global,bengana2019bifurcation}.

We emphasize that the RZIF property is not universal for oscillating
flows, since Turton {\it et al.~}\cite{turton2015} have shown that the
standing waves in thermosolutal convection emphatically do not satisfy
this property.  Nor is RZIF a prediction, since it relies on the mean
flow that must be determined by experiment or direct numerical
simulation.
%
%
In contrast to RZIF, the self-consistent model (SCM) developed by
Manti\v{c}-Lugo, Arratia \& Gallaire \cite{mantivc2014self} is predictive, or
rather, it greatly reduces the computational work required to
determine the frequency.  In the SCM, the mean flow equation is
approximated by assuming that only the leading eigenmode of the
linearized equation is responsible for creating the mean flow
distortion (the difference between the mean flow and the unstable
equilibrium).  This assumption is based on the fact that the temporal
spectrum of the flow under investigation is dominated by its
fundamental frequency.
The amplitude of the mode corresponding to the fundamental frequency
is chosen such that the growth rate of the linear problem is zero,
thus building into the solution the ``RZ'' portion of the RZIF
property.  For the cylinder wake, the results obtained by these
coupled equations match the mean flow and the nonlinear frequency
remarkably well \cite{mantivc2014self,mantivc2015self}. The SCM has
also been used to treat acoustic emissions in the compressible wake of
a cylinder \cite{fani2018computation} and the two-dimensional shear
driven cavity \cite{meliga2017harmonics}.
Other reduced-order models in which sets of modes or interactions are
omitted have been proposed and implemented for many other hydrodynamic
phenomena, notably in aeronautics and fluid mechanics
\cite{mckeon2010critical,mckeon2013experimental,
  hwang2010linear,mantic2016self,mantivc2016saturation,
  beneddine2016conditions,beneddine2017unsteady,
symon2018non,symon2019tale,yim2019self,rigas2021nonlinear,gayme2010streamwise,
thomas2014self,alizard2019restricted,gayme2019coherent,
yim2020nonlinear,rosenberg2019computing,pausch2019quasilinear}.
and in geophysics and astrophysics 
\cite{marston2016generalized,farrell2007structure,srinivasan2012zonostrophic,tobias2013direct,tobias2017three,allawala2020dimensional}.
Some of these models will be compared to RZIF and SCM in the next
section.

Here we investigate the self-consistent model for the traveling wave
branch in thermosolutal convection, for which RZIF is satisfied
\cite{turton2015}. We will demonstrate that, for this case, the
self-consistent model fails to predict the frequency or the mean flow.
Higher order terms contributing to the Reynolds stress are necessary
to reproduce the mean flow to sufficient accuracy.  Therefore, 
satisfaction of the RZIF property does not necessarily imply the
validity of the self-consistent model.


\section{RZIF and SCM Framework}
\label{sec:framework}

We present in this section the equations governing the
formalism of the RZIF (Real Zero Imaginary Frequency)
and SCM (Self-Consistent Model) approximations.
Consider a general dynamical system of the form
\begin{align}
  \pd_t U &=\mL U + \mN(U,U)
\label{eq:gendyn}\end{align}
where $\mL$ and $\mN$ are linear and bilinear operators, respectively,
and $U$ may depend on one or more spatial dimensions.  The operators
$\mL$ and $\mN$ depend on a control parameter $r$.  We assume that
\eqref{eq:gendyn} has an equilibrium (base) state $\Ub$ and
undergoes a supercritical Hopf bifurcation at a critical value $r_{\rm Hopf}$
leading to a stable limit cycle.  The base state satisfies
\begin{align}
  0 &= \mL \Ub + \mN(\Ub,\Ub) 
\label{eq:base}\end{align}
Classic linear stability analysis is derived by writing $U=\Ub+u$,
substituting into \eqref{eq:gendyn}:
\begin{align} 
	\pd_t u = \mL \Ub + \mL u + \mN(\Ub,\Ub)  + \mN(u,\Ub) + \mN(\Ub,u) + \mN(u,u) 
\end{align}
subtracting \eqref{eq:base}:
\begin{align} 
	\pd_t u &= \mL u + \mN(u,\Ub) + \mN(\Ub,u) + \mN(u,u) 
\end{align}
and neglecting the nonlinear terms $\mN(u,u)$:
\begin{align} 
	\pd_t u &= \mL u + \mN(u,\Ub) + \mN(\Ub,u)
\label{eq:LSA}\end{align}
Since \eqref{eq:LSA} is linear in $u$ and homogeneous in time, its solution is
of the form $u(t)=\exp[(\sL+i\oL)t] \ub$ with:
\begin{align}
  (\sL+i\oL) \ub &= \mL_{\Ub} \ub\\[-20pt]
\intertext{where we have defined: \vskip-10pt}
   \mL_{\Ub} &\equiv \mL + \mN(\Ub,\;\cdot\;) + \mN(\;\cdot\;,\Ub) \nonumber
\end{align}
Like $\mL$ and $\Ub$, the eigenvalue $\sL+i\oL$ depends on the parameter $r$.
When the growth rate $\sL$ crosses zero at $r=r_{\rm Hopf}$ and $\oL\neq 0$,
the base state $\Ub$ undergoes a supercritical Hopf bifurcation,
creating a new limit cycle $\ULC(t)$ satisfying 
\begin{align}
\pd_t \ULC(t) &= \mL \ULC(t) + \mN(\ULC(t),\ULC(t))
  \end{align}
and whose frequency is $\oL$ at onset.
For $r$ beyond $r_{\rm Hopf}$, the frequency $\oLC$ of
the limit cycle is no longer equal to $\oL$.

We now consider the temporal mean $\oU$ of the limit cycle $\ULC(t)$:
\begin{align}
	\oU \equiv \frac{1}{\TLC} \int^{\TLC}_{t=0} \ULC(t)\:dt
	\label{eq:meanDef}
 \end{align}
where $\TLC=2\pi/\oLC$. Substituting the Reynolds decomposition $U = \oU + u$
into the governing equations \eqref{eq:gendyn}, we obtain 
\begin{align} 
	\pd_t u  =  \mL \oU  + \mL u +  \mN(\oU,\oU)  + \mN(u,\oU) + \mN(\oU,u) + \mN(u,u) 
	\label{eq:NS'}
\end{align}
The temporal average of \eqref{eq:NS'} gives the equations obeyed by the mean fields 
\begin{align}
  0  = \mL \oU + \mN(\oU,\oU) + \overline{\mN(u,u)}
	\label{eq:NSM}
\end{align}
where the nonlinear interaction term $\overline{\mN(u,u)}$ is the force resulting from 
what is called the Reynolds stress in the context of hydrodynamics. It can also be viewed as the 
external force that would be required for the mean field to be a stationary
solution \cite{barkley2006linear}.
The mean field is computed from nonlinear
simulations because equation \eqref{eq:NSM}, unlike \eqref{eq:base}, is not a closed
system. By subtracting \eqref{eq:NSM} from \eqref{eq:NS'},
we obtain the exact fluctuation equations
\begin{align}
	\pd_t u  =  \underbrace{\mL u +  \mN(u,\oU) + \mN(\oU,u)}_{\mL_{\oU} u} + \underbrace{\mN(u,u) - \overline{\mN(u,u)} }_g
	\label{eq:LSAm}
\end{align}

\subsection{RZIF}

The RZIF procedure calls for omitting the nonlinear terms $g$ from 
equations \eqref{eq:LSAm}.
This omission is exact
if the nonlinear self interaction $\mN(u,u)$ of the deviation $u$ from the mean
contributes only to the mean.
(We will discuss this point further in section \ref{sec:harmonic}.)
%
This leaves
\begin{align} 
	\pd_t u &= \mL_{\oU} u \equiv \mL u  + \mN(u,\oU) + \mN(\oU,u)
	\label{eq:LSAmTruncat}
\end{align}
Since \eqref{eq:LSAmTruncat} is linear in $u$ and
homogeneous in $t$, its solutions are again of the form
$u(t) = \exp[(\sR + i  \oR)t] \uR$, leading again to the eigenproblem
\begin{align}
  (\sR+i\oR) \uR &= \mL_{\oU} \uR  \label{eq:LT2}
\end{align}
Limit cycles satisfy the RZIF property if the imaginary part 
$\oR$ of the leading eigenmode of $\mL_{\oU}$ is equal to 
the frequency $\oLC$ of the nonlinear limit cycle $\ULC(t)$
and the real part $\sR$ is zero.
Since $g$ in \eqref{eq:LSAm} is exactly zero only 
under special circumstances, RZIF will typically be
be satisfied only approximately. 
Equations \eqref{eq:NSM} and \eqref{eq:LSAmTruncat} comprise 
the linearization about the mean fields studied in
\cite{pier2002frequency,barkley2006linear,sipp2007global,mittal2008global,turton2015,
bengana2019bifurcation,bengana2019spirals}

\subsection{SCM}

The RZIF equations \eqref{eq:NSM} and \eqref{eq:LT2} are not
predictive or closed, because the mean flow $\oU$ must be computed in
some other way, sometimes from experimental data but more often by
time averaging the results of a full direct numerical simulation of
the limit cycle.  In contrast, the Self-Consistent Model (SCM)
developed by Manti\v{c}-Lugo $et~al.$ \cite{mantivc2014self} does not
require the mean flow as an input. Instead, these authors make the
further hypothesis that the contribution from the leading eigenmode
suffices to generate the mean flow distortion, i.e.~its deviation from
the base flow.  According to this approximation, $u$ in
$\overline{\mN(u,u)}$ in \eqref{eq:NSM} is no longer the deviation
from the mean of the limit cycle, but an eigenvector $\uS$.  Moreover,
they hypothesize that $\uS$ can be chosen (via its amplitude; see
section \ref{sec:algorithms}) such that the real part of the
eigenvalue is zero, i.e. such that $\US$ is marginally stable.  This
leads to the problem:
\begin{subequations}
\begin{align}
  0  = &\mL \US + \mN(\US,\US) +\mN(\uS,\uS^\ast)
\label{eq:NSMtrunc}\\
  i\oS \uS = &\mL_{\US} \uS  
\label{eq:LT2trunc} \\[-20pt]
\shortintertext{where \vskip-10pt}
 &\mL_{\US} \equiv \mL + \mN(\;\cdot\;,\US) + \mN(\US,\;\cdot\;)\nonumber
\end{align}
\end{subequations}
Table \ref{tab:systems} summarizes the linear stability problem
and the RZIF and SCM approximations.

\begin{table}
\begin{tabular}{|l|c|l|c|l|}
\hline
\multicolumn{2}{|c|}{Name} & Linearize about & System & Property\\
\hline
        &&&&\\    
    LSA & Linear Stability Analysis & Base flow $\Ub$ & $0 = \mL \Ub +\mN(\Ub,\Ub)$ &\\[0.1cm]
&&&                          $(\sL+i\oL)\ub = \mL_{\Ub} \ub$ &\\
        &&&&\\    \hline
        &&&&\\    
    RZIF & Real Zero & Mean flow $\oU$ &
           $\pd_t \ULC=\mL \ULC + \mN(\ULC,\ULC)$ & \\
     & Imaginary Frequency &  & $\ULC(\TLC)=\ULC(0)$ \qquad  $\oU \equiv \frac{1}{\TLC}\int_0^{\TLC} \ULC(t)\: dt$& $\oR=\oLC$
              \\[0.1cm]
&&&     $(\sR+i\oR)\uR = \mL_{\oU} \uR$ & $\sR=0$\\
        &&&&\\\hline
        &&&&\\
    SCM & Self-Consistent Model &Approximate& $0 = \mL \US +\mN(\US,\US) + \mN(\uS,\uS^\ast)$ & \\[0.1cm]
        &&mean flow $\US$  &    $i\oS\uS = \mL_{\US} \uS$  &  $\oS=\oLC$\\
    &&&&\\ \hline
\end{tabular}
\caption{Specification of classic linear stability analysis
  about the base flow (LSA), linearization about the mean (RZIF),
  and the self-consistent model (SCM). The equations in the
  column labelled System define the problem, while the
  equations in the column labelled Property may or may not be
  satisfied by the corresponding system, or may be satisfied only
  approximately.}
\label{tab:systems}
\end{table}

\subsection{Semilinear or quasilinear models}

To place RZIF and SCM in context, these are variants of a
large family of approximations based on partitioning the velocity
field into two components, $\oU$ and $u$.
$\oU$ varies, if at all, only on large spatial or temporal scales, while 
$u$ is governed by an equation that depends on $\oU$
and is linear in $u$.
The equation for $\oU$ contains nonlinear terms in $u$ which influence $\oU$; 
for the Navier-Stokes equations,
these are the quadratic terms arising from the Reynolds stress.
Nonlinear terms in $u$ which do not contribute to $\oU$ are omitted.

Such approximations can be classified according to the type of partition,
i.e. what defines the set $\oU$ and $u$. 
The RZIF and SCM approximations partition in the temporal frequency domain.
$\oU$ is the temporal mean and $u$ the time-varying field.
Since $\oU$ is the temporal mean, it is constant, 
and since $u$ satisfies a linear equation, it is an eigenvector.
These approximations are therefore not suitable for time integration.
Instead, they have been used to determine the frequency
and to approximate the spatio-temporal form of a limit cycle.

McKeon \& Sharma \cite{mckeon2010critical} proposed a temporal
partition approach in which $g\equiv\mN(u,u) - \overline{\mN(u,u)}$
in \eqref{eq:LSAm} is not
omitted but instead considered as an input to the transfer
function or resolvent operator $(i\omega - \mL_{\oU})^{-1}$.
Note that if \eqref{eq:LT2} holds with $\sigma=0$,
then $(i\omega - \mL_{\oU})$
has a non-trivial kernel and is therefore non-invertible.  In
the resolvent approach, $(i\omega - \mL_{\oU})$ is considered to be
invertible but to have one or a few singular values much smaller than
the others.  The resolvent $(i\omega - \mL_{\oU})^{-1}$ then acts as a
filter by highly amplifying the component(s) in $g$ of the
corresponding singular vector(s).
The resolvent is often studied in the context of the 
optimal forcing problem, that of determining the forcing function
and frequency which are maximally amplified. 
In its most basic form, this problem is:
\begin{subequations}
\begin{align} 
  0 &= \mL \oU + \mN(\oU,\oU) + \overline{\mN(u,u)}\label{eq:resolventa}\\
	\left(i\omega - \mL_{\oU}\right) u &= f e^{i\omega t}\label{eq:resolventb}
\end{align}
\label{eq:resolvent}
\end{subequations}
\noindent As in the distinction between RZIF and SCM, two variants are
possible: \eqref{eq:resolventb} can be solved on its own using the
exact mean $\oU$, or it can use the $\oU$ calculated self-consistently
by the coupled system \eqref{eq:resolventa}-\eqref{eq:resolventb}.
The nonlinear optimal forcing problem is more exact than the linear
optimal forcing problem, since it retains in \eqref{eq:resolventb}
the nonlinear terms $g$ defined in \eqref{eq:LSAm}
as well as the imposed forcing function $f$.
The resolvent and generalizations of it have been used in
\cite{mckeon2010critical,mckeon2013experimental,
  hwang2010linear,mantic2016self,mantivc2016saturation,
  beneddine2016conditions,beneddine2017unsteady,
  symon2018non,symon2019tale,yim2019self,rigas2021nonlinear} to
approximate the optimal forcing or the energy spectrum of complex and
even turbulent flows.

A complementary approach partitions the spatial, rather than temporal,
dependence of solutions into a spatial mean $\langle U \rangle $ and spatially varying
perturbations $u$.
These approximations are sometimes called QL (QuasiLinear) models.
Like projections of the governing equations onto a set of spatial basis
functions, they can be integrated in time in the same way as
the original equations:
\begin{subequations}
\begin{align} 
  \pd_t \langle U\rangle  &= \mL \langle U \rangle + \mN\left(\langle U\rangle,\langle U\rangle\right) + \langle\mN(u,u)\rangle\\
	\pd_t u &= \mL_{\langle U \rangle} u 
\end{align}
\label{eq:QL}
\end{subequations}
where $\langle\quad\rangle$ is a spatial average.

One example of a spatial partition is the Restricted NonLinear (RNL)
model used by fluid-dynamical researchers to study wall-bounded shear
flows, in which $\oU$ and $u$ are set to be the streamwise-averaged
and streamwise-varying modes
\cite{thomas2014self,gayme2019coherent,alizard2019restricted}.  This
model has reproduced many features of transitional and turbulent pipe
flow and plane Couette flow.  A similar approach is used in
\cite{yim2020nonlinear} to study the centrifugal instability on a
vortex.  One important current of research interprets transition to
turbulence in wall-bounded shear flows as a skeleton of trajectories
connecting steady states, traveling waves and periodic orbits, and
other low-dimensional invariant dynamical objects, called Exact
Coherent Structures (ECS) in this context.  These have been computed
using the full Navier-Stokes equations, and successfully approximated
via the RNL model in
\cite{rosenberg2019computing,pausch2019quasilinear}.

QL models have been widely used in the 
geophysical and astrophysical community.
Marston, Chini \& Tobias~\cite{marston2016generalized} have 
generalized this approach to the GQL (Generalized QuasiLinear)
approximation. In the GQL, a larger set of modes (usually those with
low wavenumber) is treated in the same way as the mean, by
including all nonlinear interactions involving this set, and excluding
nonlinear interactions within the remaining (usually high wavenumber)
modes that do not contribute to the low wavenumber set.
The QL and GQL approximations have been used
to calculate the east-west bands or jets on planetary surfaces
\cite{farrell2007structure,srinivasan2012zonostrophic,tobias2013direct,
marston2016generalized}.
A study of rotating plane Couette flow \cite{tobias2017three} has provided an 
illustration of the ability of GQL to capture
features that QL does not.
Another type of mean flow for which the quasilinear approach 
can be used is the ensemble average \cite{allawala2020dimensional}.
Ensemble averaging, like temporal averaging, can also be combined
with averaging over a homogeneous spatial direction as in 
\cite{mckeon2010critical,rigas2021nonlinear}. 

Neither RZIF nor SCM fall precisely into the category of QL or GQL methods; 
see section \ref{sec:higher-order}.

\section{Application to thermosolutal waves}
\label{sec:applic}

We now turn to the hydrodynamic system for which we will compare RZIF and SCM.
A density gradient in a fluid layer often leads to convection,
i.e.~overturning motion that tends to equalize the density in the bulk.
The density gradient is in turn usually the consequence of thermal and/or concentration
gradients; when both are present, terms such as thermosolutal,
double-diffusive, and binary are used for different variants of the problem.
If the thermal and solutal effects oppose one another,
then convection can take the form of time-dependent solutions.

The thermosolutal problem studied here and in
\cite{turton2015,tuckerman2001thermosolutal} is formulated in an
idealized two-dimensional horizontally periodic domain 
$(x,z)\in[0,2.8)\times[0,1]$, 
allowing the velocity to be represented as $\nabla\times\Psi\bey$
and the equations to be stated in streamfunction-vorticity form.
At the top and bottom boundaries
$z=0,1$, different values are imposed for the temperature and
concentration, and free-slip conditions are imposed on the velocity.
There exists a motionless conductive
solution in which the temperature and concentration fields 
are linear functions of the vertical coordinate $z$.
We set $\Theta$ and $C$ to be deviations of the
temperature and concentration fields from the conductive profiles. 

The nondimensionalized governing equations are:
\begin{subequations}
\begin{align}
\pd_t \Temp - \mJ[\Psi,\Temp] &= \lap \Temp + \pd_x \Psi\\
\pd_t C - \mJ[\Psi,C] &= L\lap C + \pd_x \Psi\\
\label{eq:goveqvort}
\pd_t \lap \Psi - \mJ[\Psi, \lap\Psi]  &=P\left(\lapd \Psi + R_T \pd_x \left(\Temp+S C\right)\right)
\end{align}
\label{eq:goveq}
\end{subequations}
where the Poisson bracket is
\begin{align}
\mJ[f,g] &\equiv \bey\cdot\nabla f\times\nabla g = \pd_zf \pd_xg -\pd_xf \pd_zg 
\label{eq:Poisson}\end{align}
The ratio $P$ of kinematic viscosity to thermal diffusivity is fixed at 10
and the ratio $L$ of solutal to thermal diffusivity to 0.1 (these are the usual Prandtl
number and inverse of the Lewis number, respectively).
The imposed concentration and thermal gradients both contribute to the density
gradient and the ratio $S$ of their contributions is fixed at $-0.5$.
We vary the imposed thermal gradient, which is given in terms of the reduced Rayleigh
number $r$, the ratio of the Rayleigh number $R_T$ to its critical
value 657.5 for this geometry and in the absence of a concentration gradient.
The conductive solution is stable for $r$ until
$r=2.05$, when a Hopf bifurcation breaks the
translational symmetry in this periodic geometry, leading to the
creation of branches of traveling and standing waves
\cite{knobloch1986oscillatory}. We carry out our study over the
range $r \in[2.06,3]$.

\begin{figure}[!h]
\vspace*{-0.2cm}
\hspace*{-2cm}  \includegraphics[width=1.2\columnwidth]{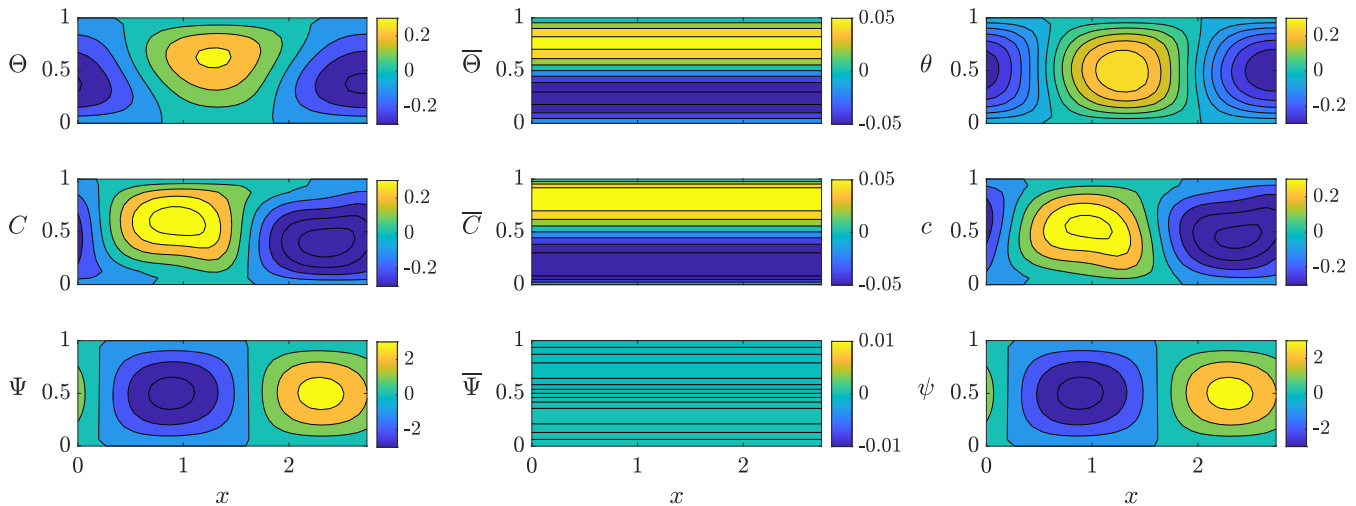}
\vspace*{-0.5cm}
\caption{Left: instantaneous snapshot $(\Theta, C, \Psi)$ of a
  traveling wave.  $\Theta$ and $C$ are the deviation of the
  temperature and concentration from the linear conductive solution,
  and $\Psi$ is the streamfunction representing the velocity.
  Center: temporal mean flow
  $(\overline{\Theta},\overline{C},\overline{\Psi})$.  Right:
  deviation $(\theta,c,\psi)$ from the mean flow.  The mean field
  $(\overline{\Theta},\overline{C},\overline{\Psi})$ is much smaller
  than the instantaneous field, so the deviation $(\theta,c,\psi)$ is
  very close to the instantaneous field.}
  \label{fig:decomp}
  \end{figure}

Figure \ref{fig:decomp} shows an instantaneous visualisation in the
$(x,z)$ plane of the exact nonlinear traveling wave $\ULC$ and its
decomposition into the temporal mean flow $\oU$ and deviation
$\ULC-\oU$. We emphasize that $\oU$ is not the conductive solution, but the
mean of the deviation from it, sometimes called the distortion.
Because $\ULC$ is a traveling wave, fields at other instants in time can
be obtained by a shift in the periodic direction $x$, and
the temporal mean is also the spatial mean in $x$.
A detailed study of the RZIF property in traveling and standing waves
in thermosolutal convection was carried out in \cite{turton2015}.
Turton et al.~\cite{turton2015} showed that the 
traveling waves had the RZIF property while standing waves at the same
parameter values did not. We will extend the study of the 
thermosolutal traveling waves to the SCM approximation.  We do not
include the standing waves, since the SCM approximation presupposes the
validity of the RZIF approximation.

\begin{figure}[t!]
  \begin{minipage}[c]{0.5\linewidth}
    \centerline{\includegraphics[width=0.9\linewidth]{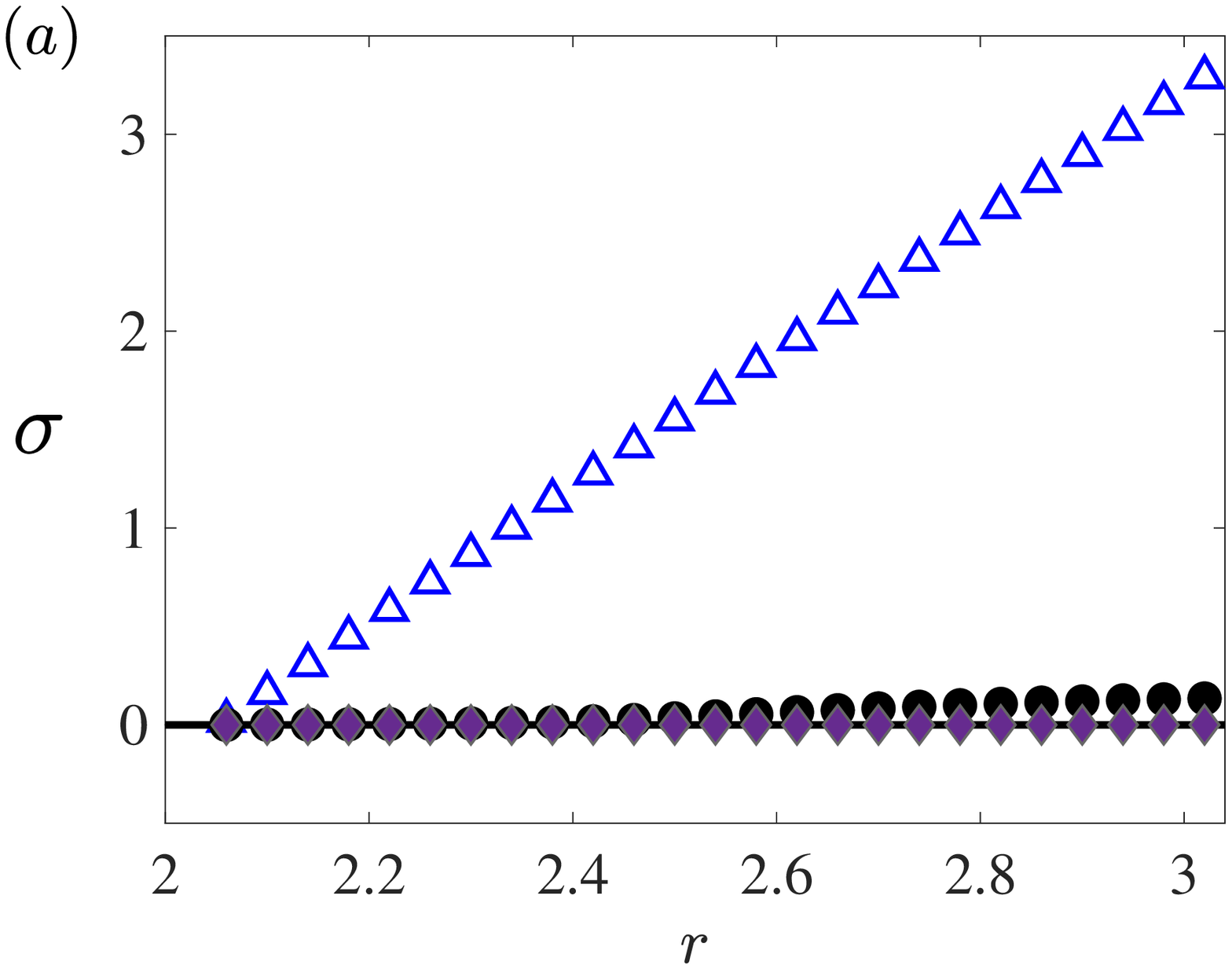}}
  \end{minipage}\hfill
  \begin{minipage}[c]{0.5\linewidth}
    \centerline{\includegraphics[width=0.9\linewidth]{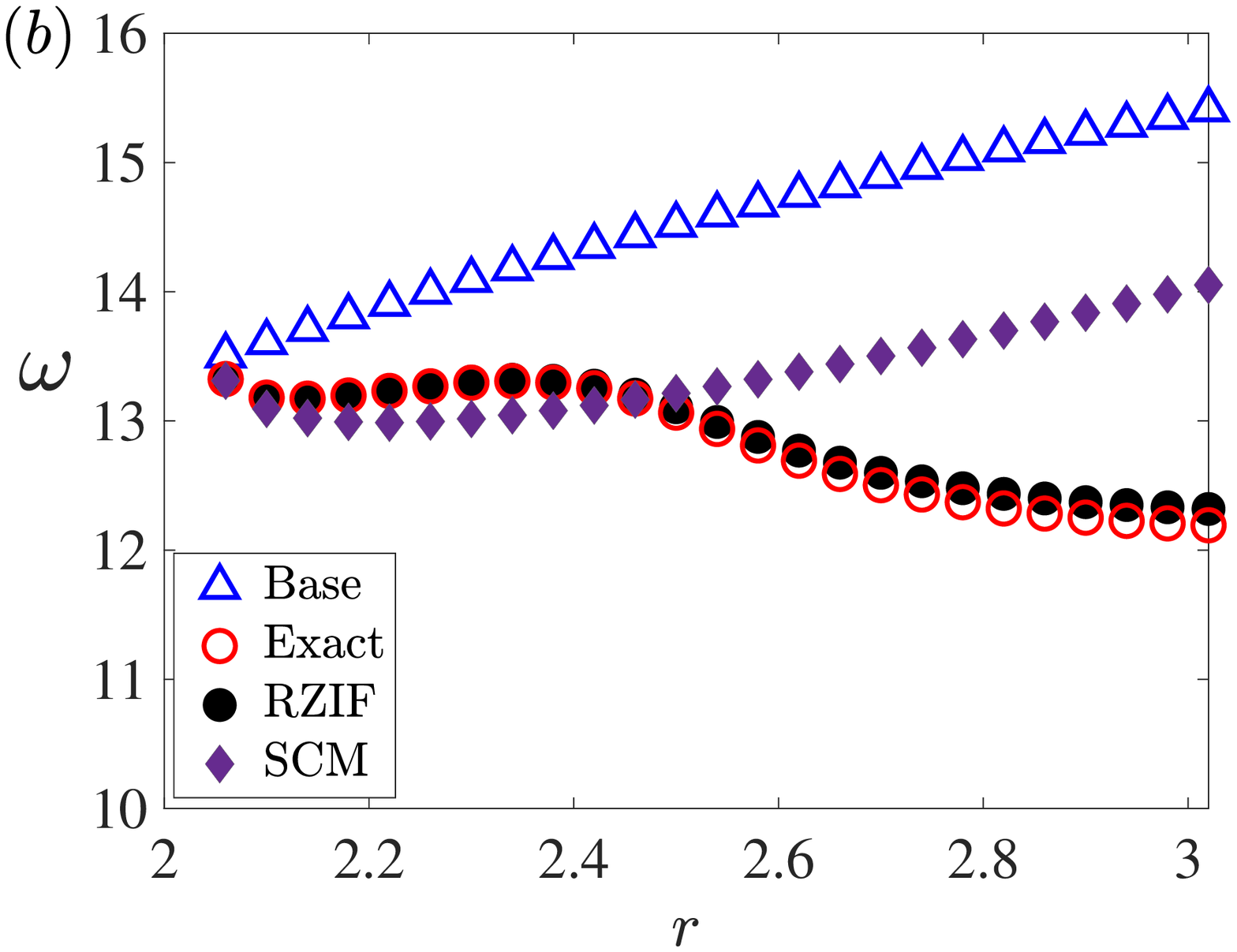}}
  \end{minipage}
  \caption{(a) Growth rate and (b) frequency as a function of Rayleigh
    number. Exact frequencies are shown by open circles
    ($\color[rgb]{1.,0.,0.}\circ$). Frequencies and growth rates
    obtained by linearization about the conductive base state are
    represented by triangles ($\color[rgb]{0.,0.,1.}\bigtriangleup$)
    while those obtained by linearization about the full mean field
    (RZIF procedure) are represented by solid circles ($\bullet$).
    Frequencies obtained by the SCM procedure are shown by diamonds
    (${\color[rgb]{0.5,0.,0.5}\blacklozenge}$). }
\label{fig:ScmOmg1st}
\end{figure} 

The main result of this study is contained in 
figure \ref{fig:ScmOmg1st}, which shows the exact frequency $\oLC$ of the
limit cycle, along with the real and imaginary parts of
the eigenvalues of the operators $\mL_{\Ub}$, $\mL_{\oU}$, and
$\mL_{\US}$ as a function of $r$.
The frequency $\oL$ obtained from linear stability analysis about the
conductive base state is far
from the frequency $\oLC$ of the limit cycle, as expected, while the frequency
$\oR$ obtained by RZIF, i.e.~linearizing around the mean flow $\oU$,
is quite close to the exact nonlinear frequency and $\sR$ remains
small in the entire range investigated, $[2.05,3]$. In contrast, the
frequency $\oS$ obtained by SCM matches $\oLC$ only very close to the
threshold, approximately for $r\in [2.05,2.08]$ and deviates below it
for $r\geq 2.1$. However, as $r$ is increased further, the $\oS$ curve
approaches the $\oLC$ curve, crossing it at $r=2.5$ and then exceeding
it substantially.  For $r\geq 2.5$, the RZIF growth rate $\sR$ becomes
slightly positive.  The SCM growth rate $\sS$ is zero for all $r$ by construction.

\begin{figure}
\includegraphics[width=3.8cm]{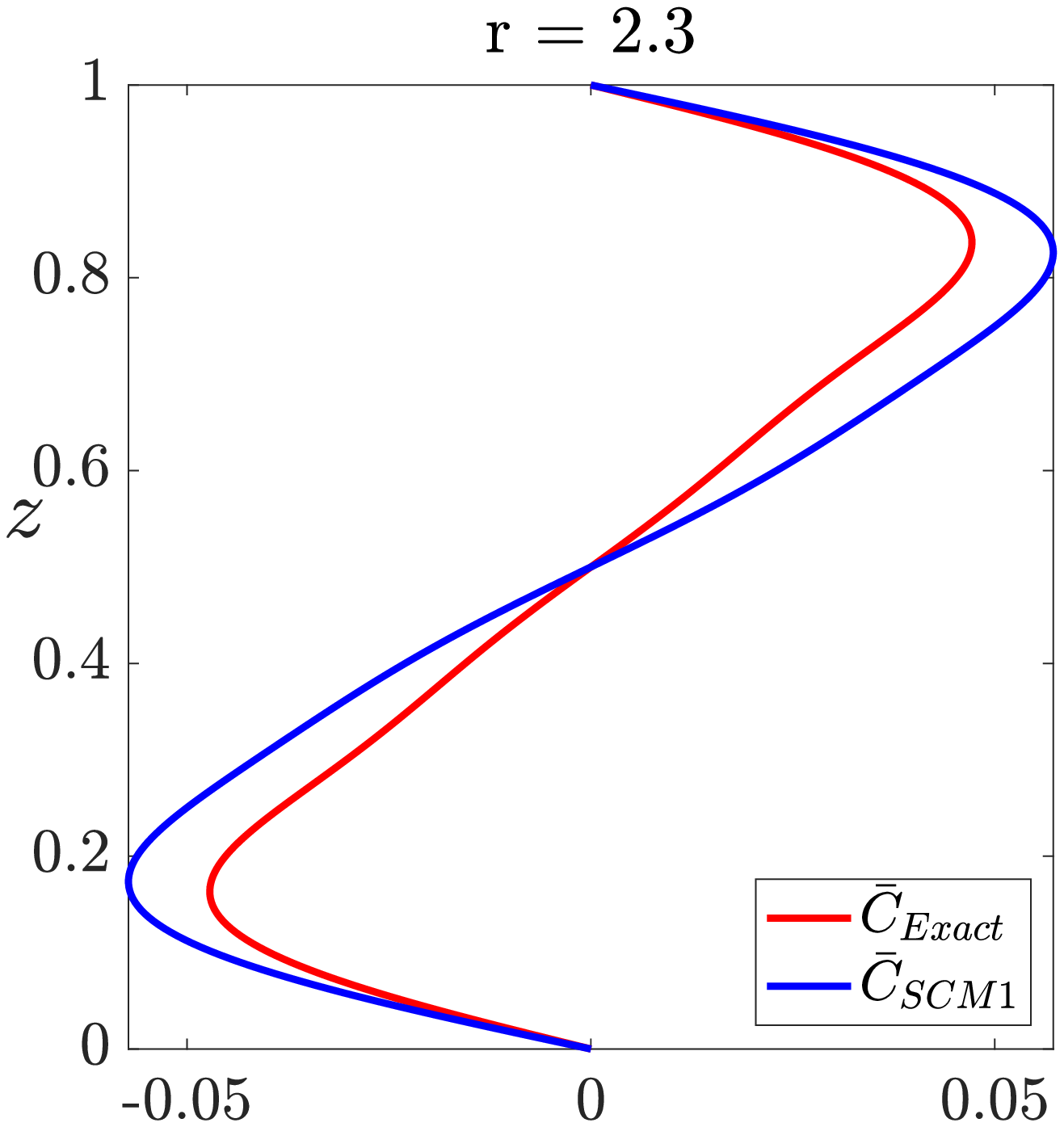}\qquad
\includegraphics[width=3.9cm]{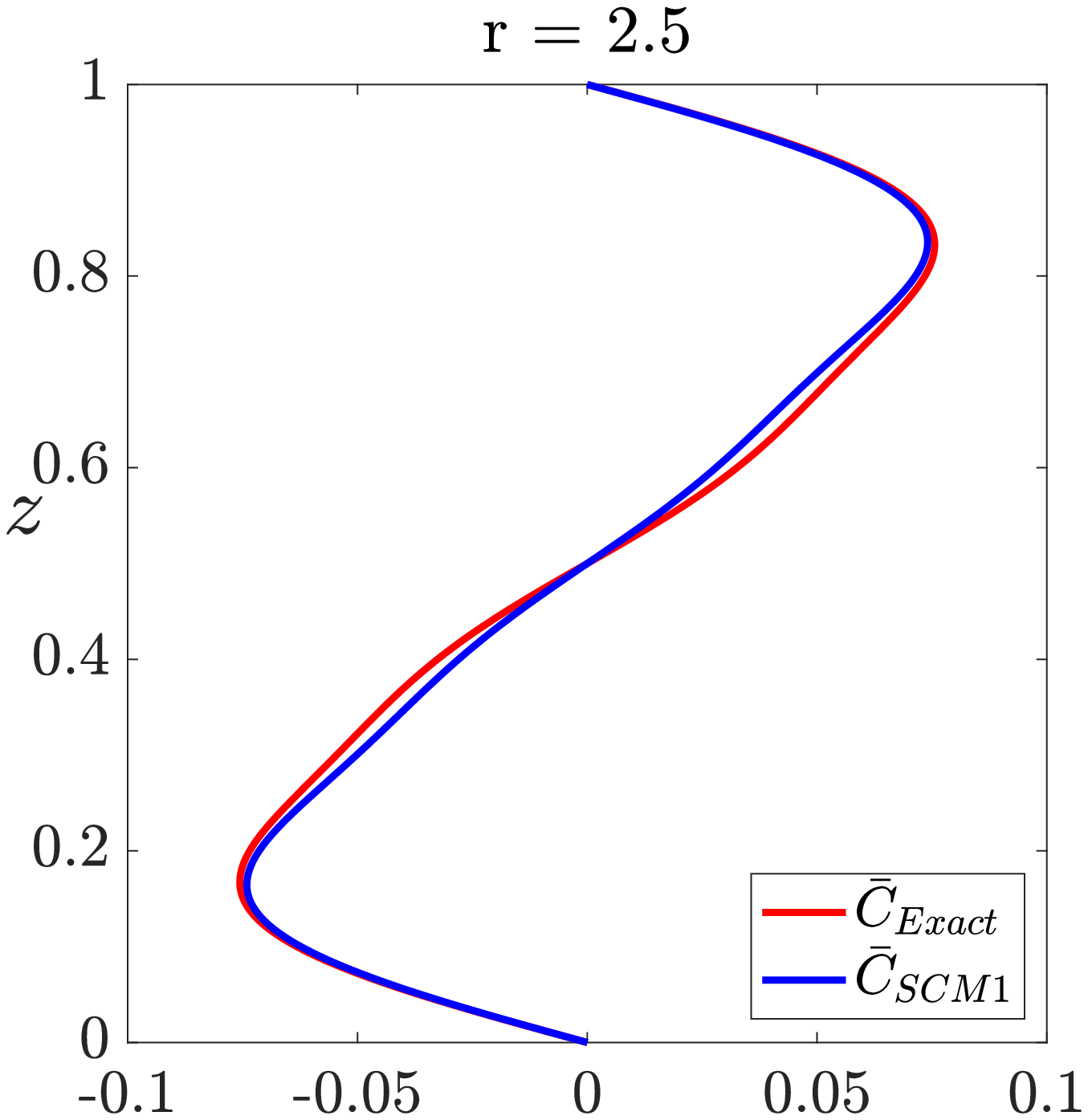}\qquad
\includegraphics[width=3.9cm]{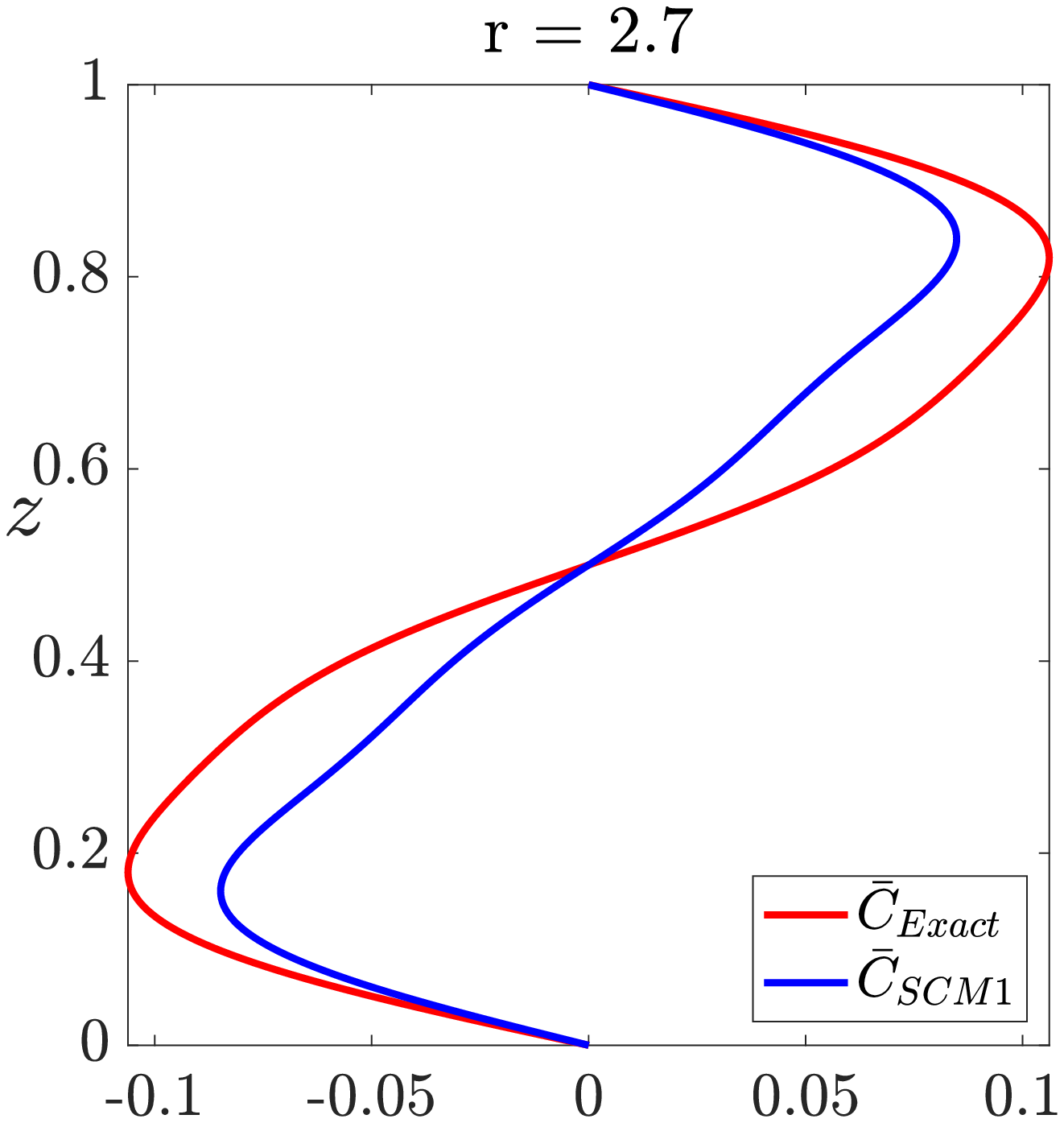}\qquad
\includegraphics[width=4.2cm]{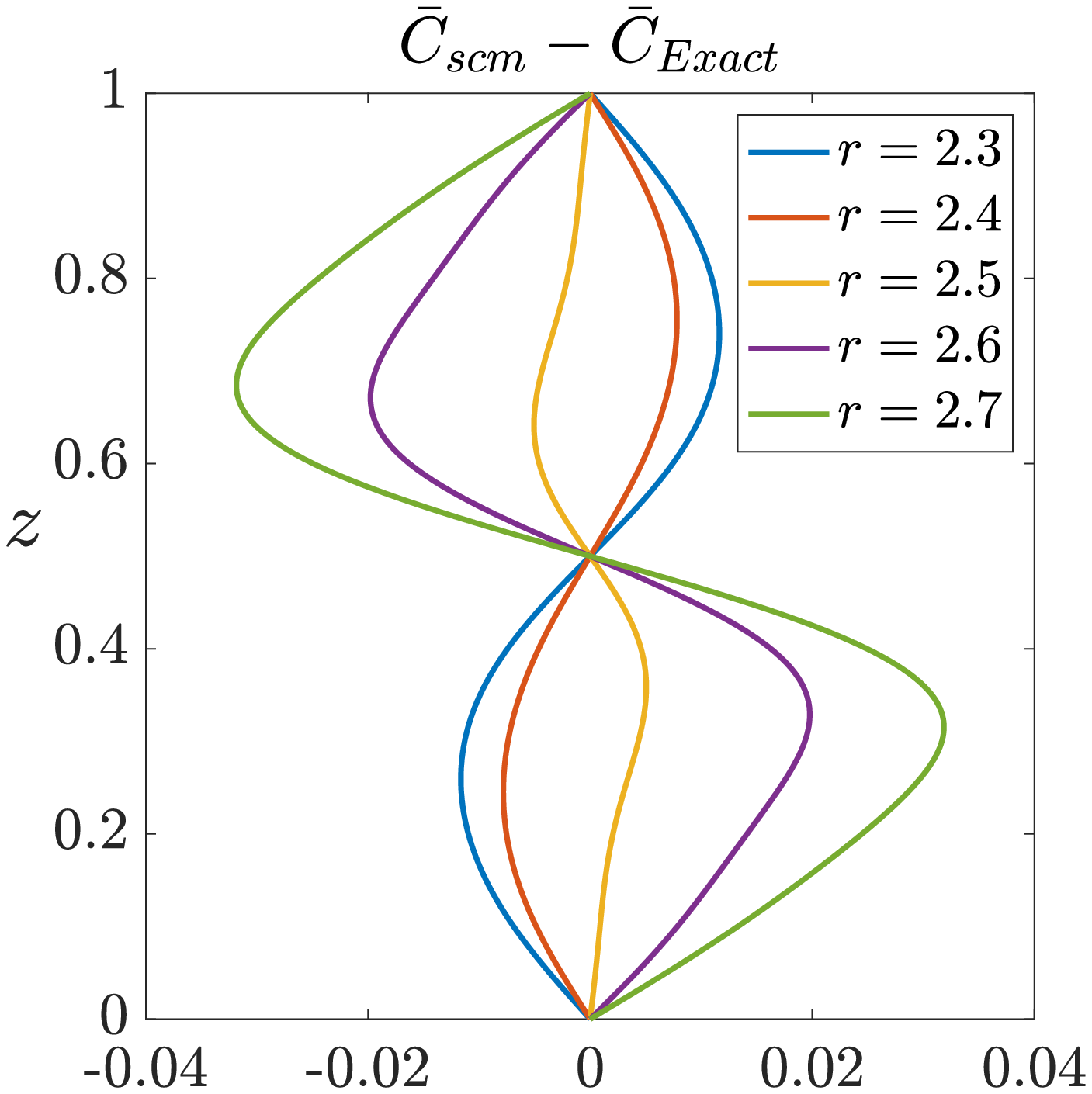}\\
\vspace*{-3.8cm}
{\it \hspace*{-4.3cm}(a) \hspace{3.9cm}(b)\hspace{4cm} (c)\hspace{4.1cm} (d)}
\vspace*{4cm}
      \caption{Mean concentration profile calculated via SCM compared to
        the exact profile. (a) For $r=2.3$, the amplitude of $\overline{C}_{\rm SCM}$ is greater than that of $\overline{C}_{\rm exact}$. (b) For $r=2.5$, the value at which $\oS \approx \oLC$, the two profiles are almost identical.
        (c) For $r=2.7$, the amplitude of $\overline{C}_{\rm scm}$ is less than that of $\overline{C}_{\rm Exact}$. (d) Difference $\overline{C}_{\rm scm}-\overline{C}_{\rm Exact}$
      for $2.3 \leq r \leq 2.7$.}
\label{fig:C_exactSCMmeanprofiles}
\end{figure}

In figure \ref{fig:C_exactSCMmeanprofiles}, we compare the mean
concentration profile calculated by the SCM to the exact mean profile
for various values of $r$.
(Recall that the RZIF procedure uses the exact mean profile.)
We choose the concentration, here and elsewhere, because the differences
are largest for this component.
The disagreement between the SCM and exact
profiles closely follows the tendency seen in figure \ref{fig:ScmOmg1st}:
a disagreement at $r=2.3$, which decreases to the point of being almost
negligible at $r=2.5$ and then increases again with $r$.
We also note that the sign of the error in the mean flows reverses at $r=2.5$, 
just as was seen for the frequency in figure \ref{fig:ScmOmg1st}.
Thus, the crossing of $\oS$ and $\oLC$ seen in figure \ref{fig:ScmOmg1st}
at $r=2.5$ is not a coincidence, e.g. two different operators sharing the 
same eigenvalues.
The agreement between the eigenvalues at $r=2.5$ 
is due precisely to the fact that the SCM approximation to the mean field
is accurate at that particular value.

This case provides a counterexample to the SCM, showing that the RZIF
property does not necessarily imply the validity of SCM. The
assumption that only the leading mode contributes significantly
to the distortion of the mean field does not hold.

Manti\v{c}-Lugo \& Gallaire \cite{mantivc2016saturation} have 
carried out a study of the optimal forcing response in the backward
facing step, comparing fully nonlinear results
(retaining in \eqref{eq:resolventb} the nonlinear terms $g$ of \eqref{eq:LSAm})
with linear results from the resolvent \eqref{eq:resolventb}, either
computed from the exact mean flow or from a self-consistent
approximation to the mean using a single mode as in \eqref{eq:resolventa}.
Surprisingly, they find that the results from the single-mode
approximation to the mean and resolvent (comparable to SCM)
are much closer to the nonlinear results 
than those using the exact mean and resolvent (comparable to RZIF).
This could be due to the consistency of the
truncation used in SCM, or to some difference between
limit cycles and optimal forcing, or between the
thermosolutal problem and the backward-facing step, 
or merely to chance.

\clearpage
\section{Fourier Analysis: Harmonic Balance}
\label{sec:harmonic}

\begin{figure}
      \includegraphics[width=\textwidth]{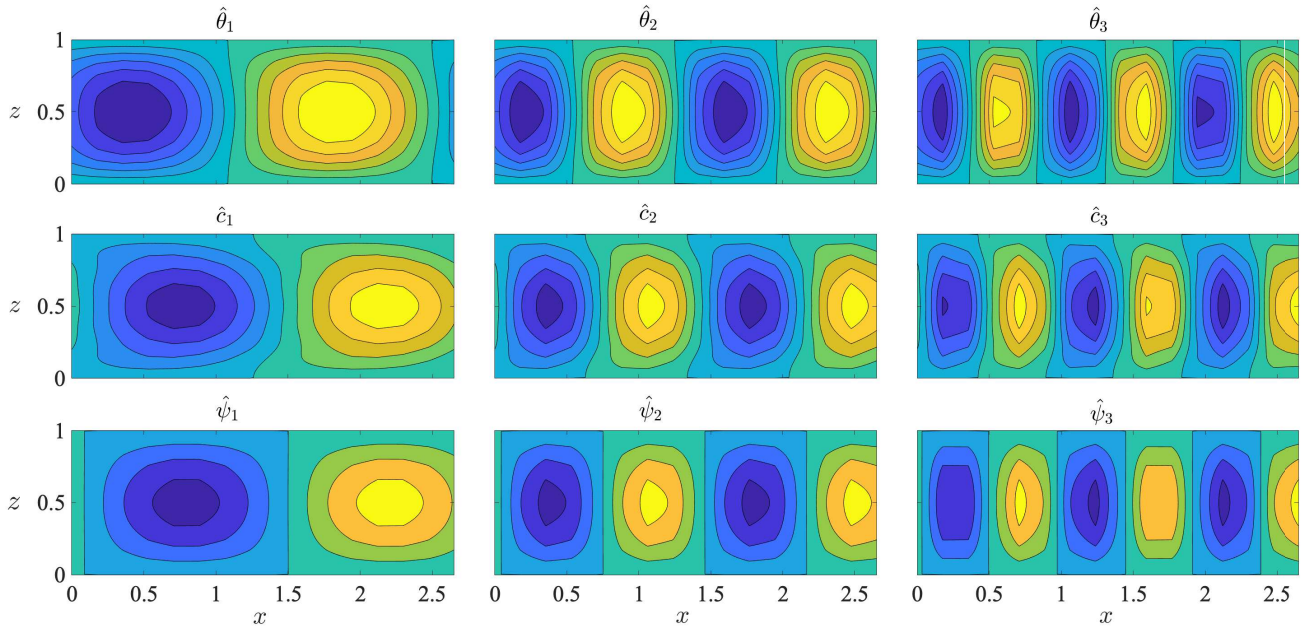}
\vspace*{-0.25cm}
      \caption{Temporal Fourier components 1, 2, and 3 for traveling wave solution
        of thermosolutal convection at $r=2.5$.
        Fourier components of temperature $\hat{\theta}$, concentration $\hat{c}$ and streamfunction $\hat{\psi}$ are shown. The components are complex, with combinations of real and imaginary part parametrized by a phase. Here, a single choice of temporal or spatial phase is shown.}
\label{fig:Fourier}
\end{figure}

To further understand the RZIF and SCM equations, we turn to the
temporal Fourier decomposition of the limit cycle and of the governing
equations.  The statement of the governing equations in terms of the
temporal Fourier decomposition is called harmonic balance in the
aerodynamic literature
\citep{rigas2021nonlinear,hall2002computation,mcmullen2006demonstration,mcmullen2006computational}
and it is the basis of the argument presented in Turton {\it et al.}
\cite{turton2015}.
We write the limit cycle $\ULC(t)$ as 
\begin{align}
    \ULC =\oU +\sum_{n\neq0}\hu_n  e^{in\omega t} 
    \label{eq:DecompFourier}
\end{align}
where $\hu_{-n}=\hu^\ast_n$. 
Figure \ref{fig:Fourier} shows these Fourier components 
for our case of traveling waves in thermosolutal convection.
Their spatial form is dictated by the fact that 
a temporal Fourier decomposition is equivalent
to a horizontal spatial Fourier decomposition for a traveling wave.

We then substitute
\eqref{eq:DecompFourier} into the governing equations \eqref{eq:gendyn}
and separate the resulting terms of different frequencies.
The term corresponding to $n=0$ is the governing equation of the mean field:
\begin{subequations}
\begin{align}
    0 = \mL\oU + \mN(\oU,\oU) + \underbrace{ \sum_{m\neq0}\mN(\hu_m,\hu_{-m})}_{\mN_0}
    \label{eq:zerothFreq}
\end{align}
\label{eq:allFreq}
The nonlinear term $\mN_0$ appearing in \eqref{eq:zerothFreq} is
the divergence of the Reynolds stress, responsible for the distortion and
production of the mean field. 
The equation corresponding to each $n>0$ is:
\begin{align}
    in\omega \hu_n&= \underbrace{ \mL\hu_n + \mN(\oU,\hu_n) +\mN(\hu_n,\oU)}_{\mL_{\oU}\hu_n} +\underbrace{ \sum_{m\neq 0,n}\mN(\hu_m,\hu_{n-m})}_{\mN_n}
    \label{eq:otherFreq}
\end{align}
\end{subequations}
For $n=1$, \eqref{eq:otherFreq} becomes
\begin{align}
    i\omega \hu_1&= \mL_{\oU}\hu_1 +\mN_1
    \label{eq:oneFreq}\\[-20pt]
\intertext{where \vskip-10pt}
  \mN_1&\equiv\mN(\hu_2,\hu_{-1})+\mN(\hu_{-1},\hu_2)+\mN(\hu_3,\hu_{-2})+\mN(\hu_{-2},\hu_3)+ \dots \nonumber
  \end{align}
  If the periodic cycle is exactly monochromatic, i.e.~if $\hu_{\pm 2}=\hu_{\pm 3}=\ldots=0$,
  then $\mN_1=0$ and \eqref{eq:oneFreq} becomes the RZIF equation \eqref{eq:LT2} with $\sR=0$:
\begin{align}
    i\omega \hu_1= \mL_{\oU}\hu_1 
\label{eq:rzifexact}\end{align}
  If, as is more likely, $u_{n \geq 2}$ is not zero, but is small,
  for example if 
\begin{align}
  ||\hu_n|| \sim \epsilon^{|n|}
\label{eq:ordereps}\end{align}
as discussed in \cite{duvsek1994numerical}, 
then $\mN_1$ is of order $\epsilon^3$, while $i\omega \hu_1$ and
$\mL_{\oU}$ are of order $\epsilon$, so that \eqref{eq:rzifexact} is approximately true.  
(Note that \eqref{eq:ordereps} does not justify neglecting $\mN_n$ in
\eqref{eq:otherFreq} for $n>1$, since 
$in\omega \hu_n$, $\mL_{\oU} \hu_n$, and $\mN_n$ are all of order $\epsilon^n$.)

The argument in terms of spectra is supported by the results of Turton et
al.~\cite{turton2015}.
We recall that standing waves are produced
at the same Hopf bifurcation as the traveling waves and that 
the RZIF property does not hold for the standing waves. In \cite{turton2015},
it is shown that 
the spectrum of the standing waves is far less peaked at $n=1$ than
that of the traveling waves. For example, at $r=2.5$,
the ratio of the Fourier components of the temperature field $||\hat{\theta}_2||/||\hat{\theta}_1||$
is approximately $10^{-2}$ for the traveling
waves and 20 times higher for the standing waves.

To be consistent, the quantitative argument based on \eqref{eq:ordereps} would also call for
neglecting terms $\mN(\hu_m,\hu_{-m})$ for $m\geq 2$ compared to $\mN(\hu_1,\hu_{-1})$,
leading to the SCM. 
The Fourier interpretation of the SCM is that the limit cycle is represented
by a temporal Fourier series, truncated to contain only modes 0 ($\oU$) and 1 ($\hu_1$).

\begin{figure}
\vspace*{-0.25cm}
\hspace*{-1cm}\includegraphics[width=0.52\columnwidth,clip]{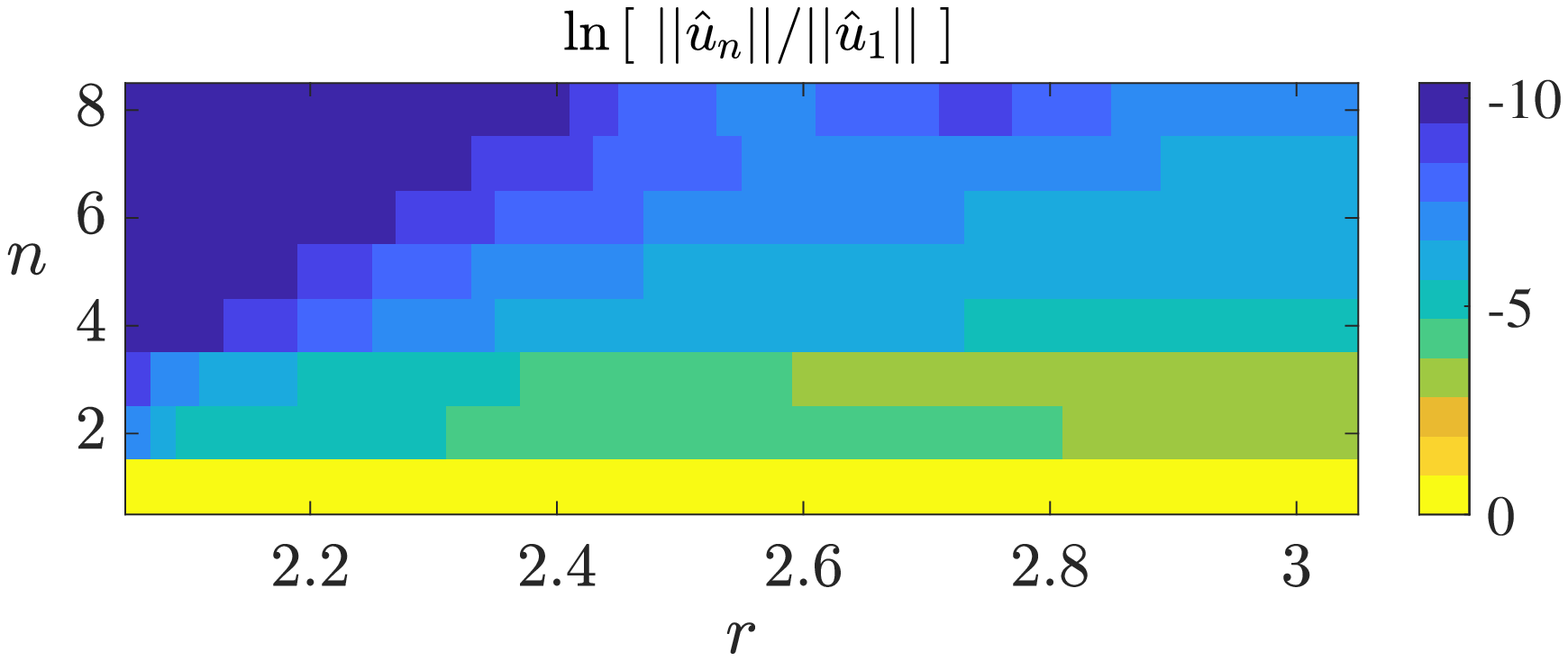}
\hspace*{-0.5cm}\includegraphics[width=0.52\columnwidth,clip]{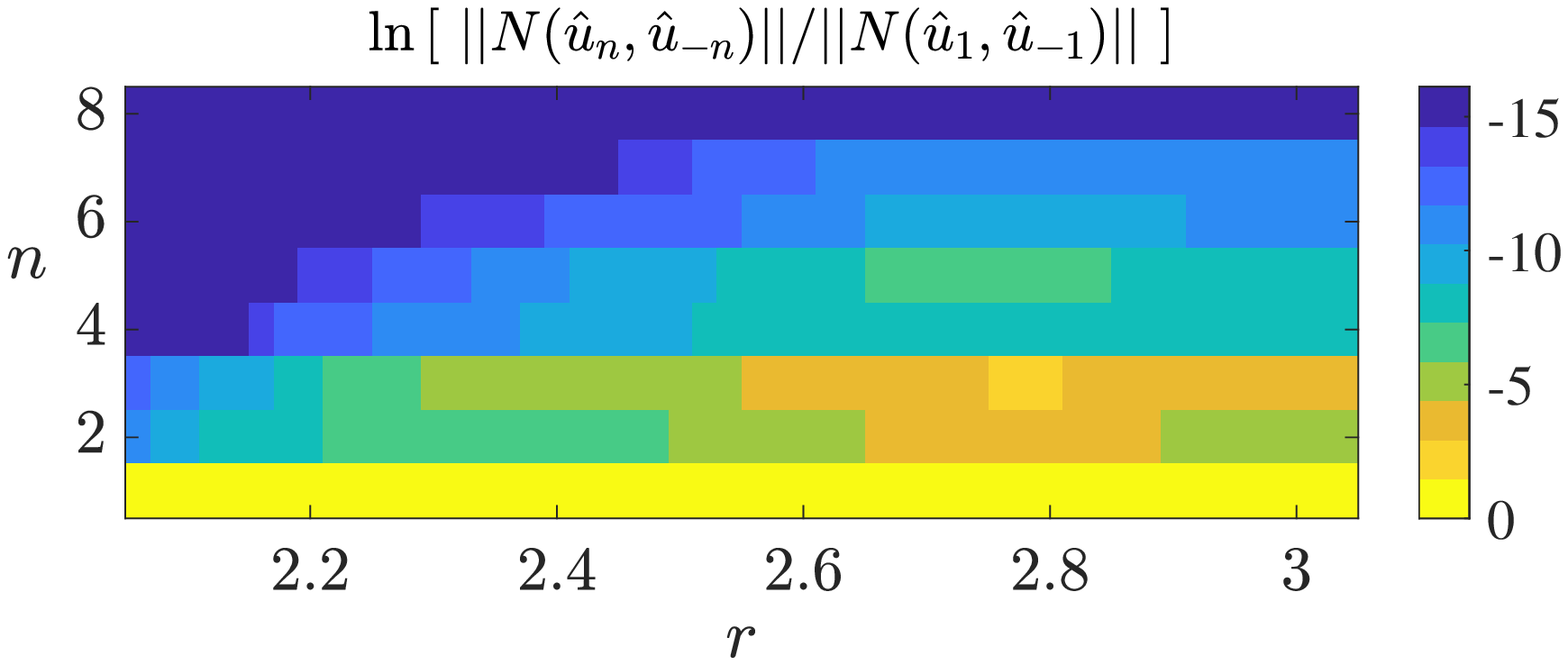}
\vspace*{-0.25cm}
      \caption{Logarithmic color representation of Fourier spectra. 
        Left: $||\hu_n||$ normalized by $||\hu_1||$. Right:  Contributions $||\mN(\hu_{-n},\hu_n)||$ to the mean flow normalized by $||\mN(\hu_{-1},\hu_1)||$.}
\label{fig:spectra}
\end{figure}

In figure \ref{fig:spectra}(a) we visualize the temporal Fourier spectra $||\hu_n||$
over the range $r\in[2.05,3]$ and for frequencies $n\in[1,8]$. We normalize by 
$||\hu_1||$ since the RZIF approximation relies on neglecting $\hu_{n>1}$
in comparison with $\hu_1$.  Figure \ref{fig:spectra}(b) 
shows the amplitudes of the nonlinear terms contributing to the mean flow
$||\mN(\hu_n,\hu_{-n})||$.
We normalize by $||\mN(\hu_1,\hu_{-1})||$, since the SCM assumes that 
$\mN(\hu_n,\hu_{-n})$ can be neglected in comparison with $\mN(\hu_1,\hu_{-1})||$.
These figures show that both spectra are highly peaked for small $r$
and become less so as $r$ increases, as is to be expected.
Going from $n=2$ to 3, the magnitudes decrease very little,
and even increase for higher values of $r$, 
a point that will be explored further in the next section.

\begin{figure}
\begin{minipage}{12.5cm}
\includegraphics[width=4.1cm]{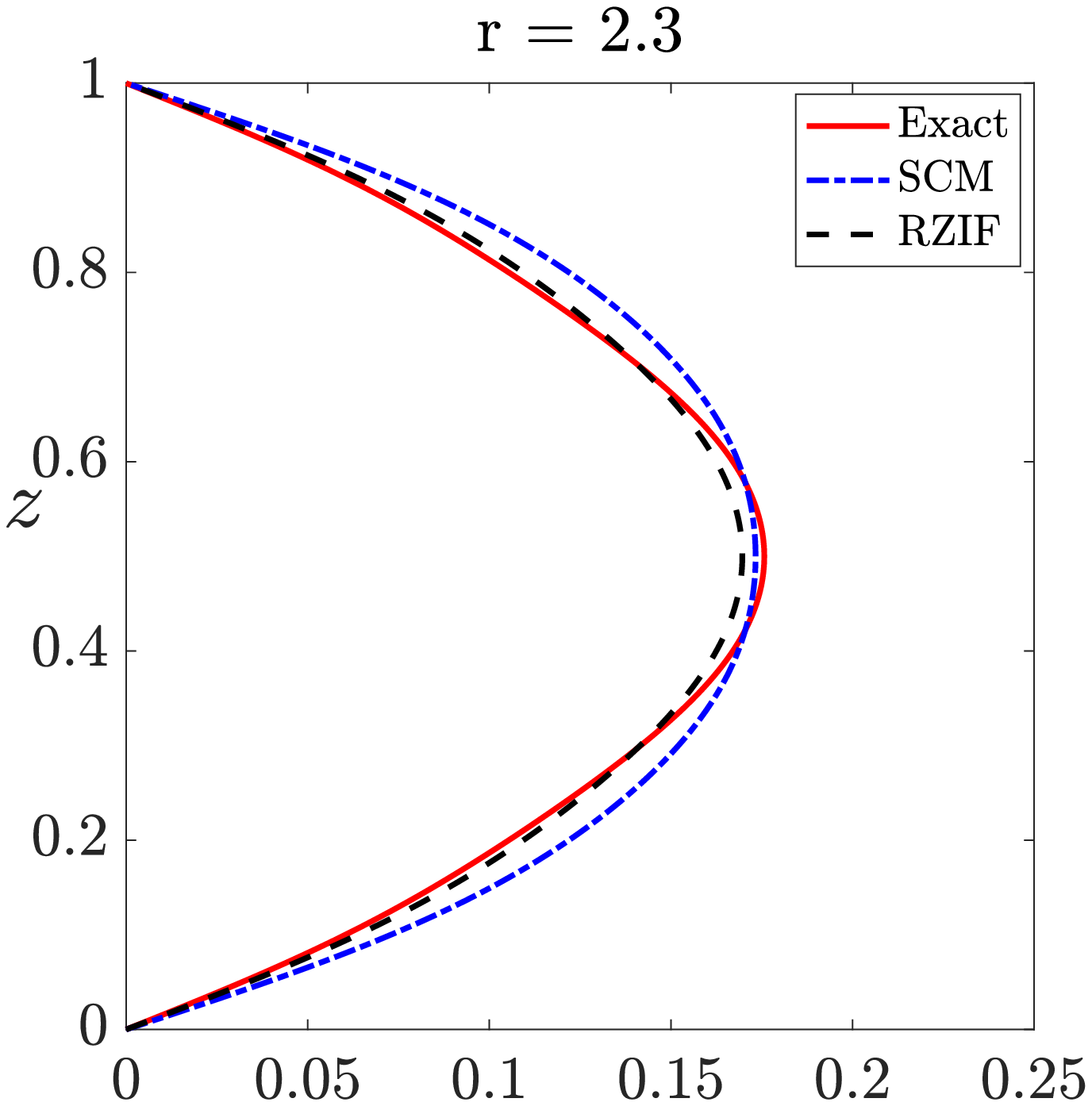}
\includegraphics[width=4.1cm]{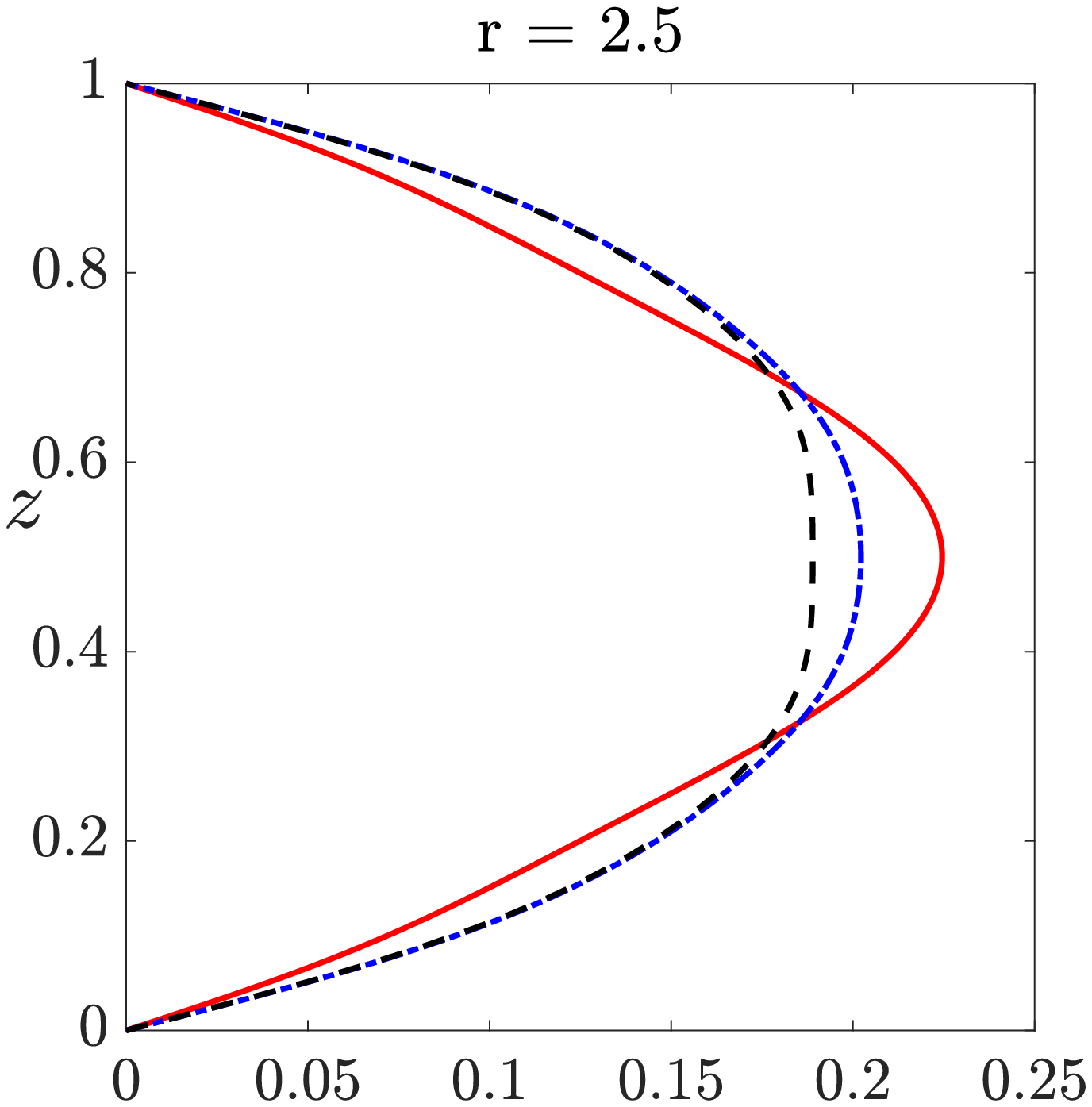}
\includegraphics[width=4.1cm]{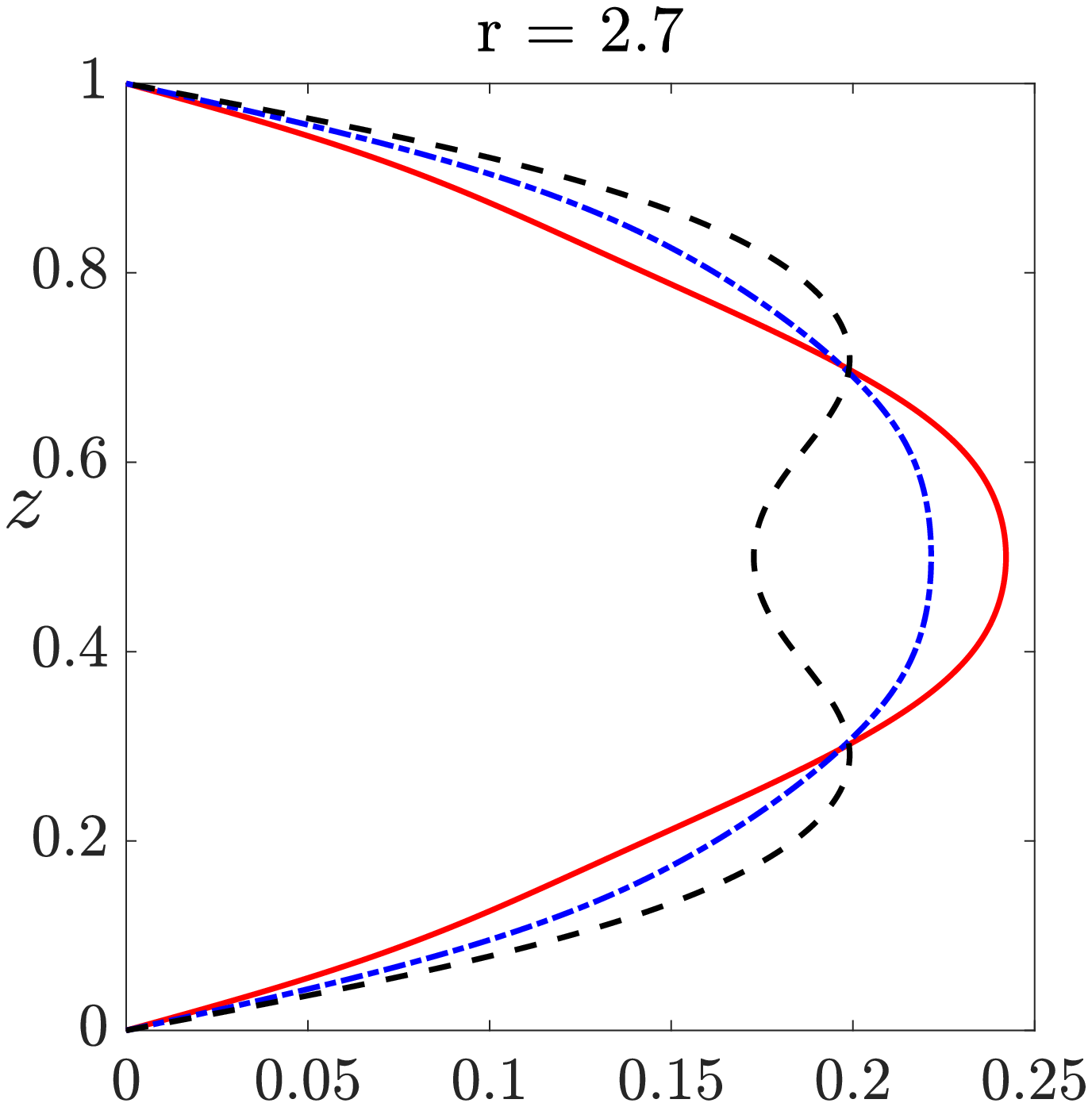}
\end{minipage}
\begin{minipage}{5cm}
\vspace*{0.5cm}
\includegraphics[width=4.9cm]{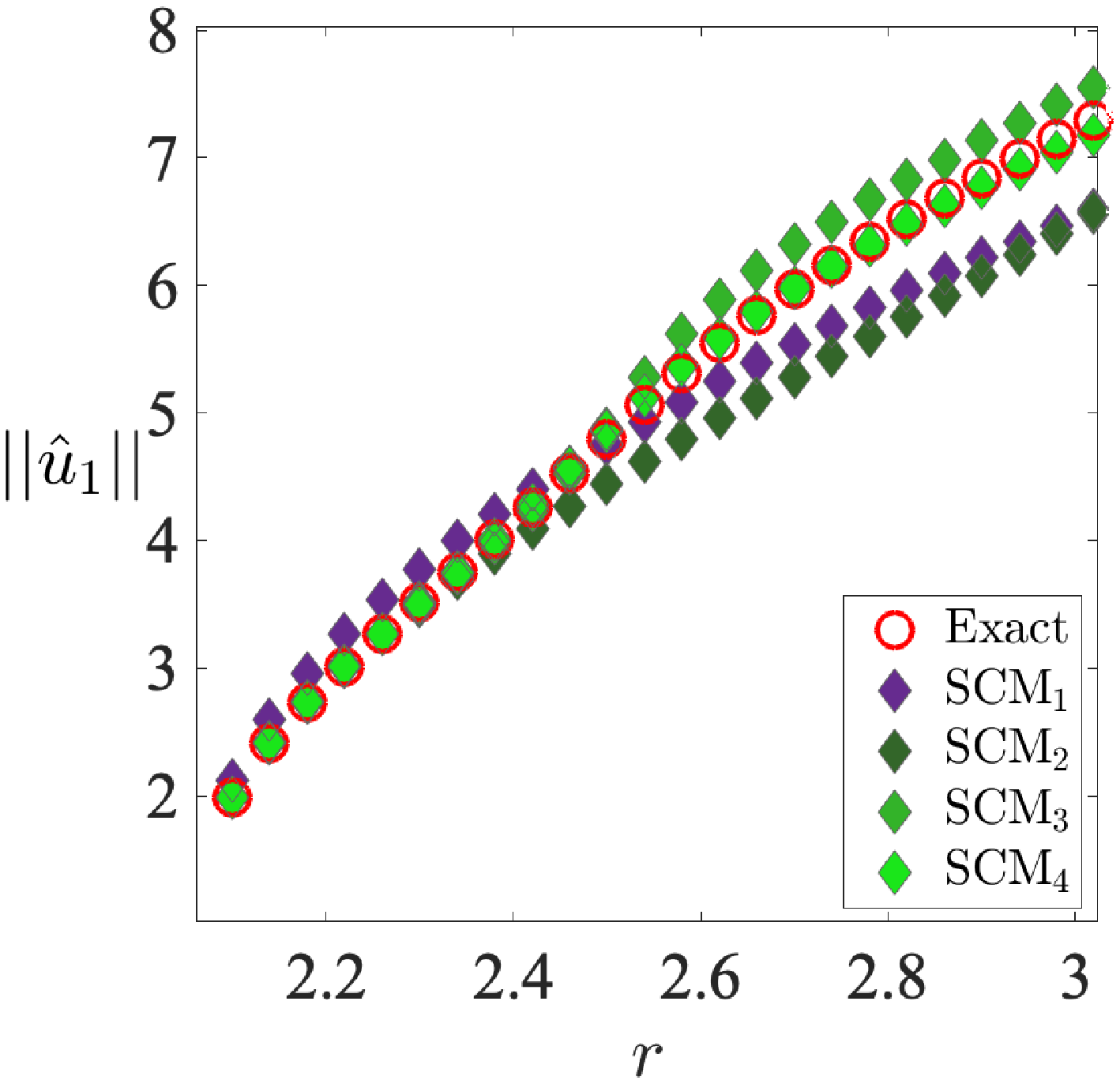}
\end{minipage}

\vspace*{-4.5cm}

{\it \hspace*{-5cm}(a) \hspace{3.7cm}(b)\hspace{3.7cm} (c)\hspace{3.7cm} (d)}
\vspace*{4cm}
\caption{Modulus of first Fourier component $|\hat{c}_1(z)|$ of concentration field 
  and its approximations via SCM and RZIF for (a) $r=2.3$, (b) $r=2.5$, and (c) $r=2.7$.
  The amplitudes of the RZIF profiles are undetermined, since they are eigenvectors; 
  here, they have been normalized to match the norms of the Fourier components.
SCM overestimates $|\hat{c}_1|$ for $r<2.5$ and underestimates it for $r>2.5$.
  (d) Norm $||\hat{u}_1||$ 
  and its approximation via various orders of SCM as a function of $r$.}
\label{fig:C_exactSCMfirstFourprofiles}
\end{figure}

According to \eqref{eq:rzifexact}, 
the RZIF procedure does not merely approximate the nonlinear
frequency as the leading eigenvalue but also approximates 
the first temporal Fourier component via the corresponding eigenvector.
Figure \ref{fig:C_exactSCMfirstFourprofiles}(a,b,c) illustrates this 
idea by comparing $|\hat{c}_1(z)|$ with its approximations via RZIF and SCM.
Since $c_{\rm rzif}$ is part of an eigenvector, its norm has been chosen to
match that of $\hat{c}_1$, i.e.~$\int dz \; |c_{\rm rzif}(z)| = \int dz \;|\hat{c}_1(z)|$.
For $r<2.5$, the SCM profile slightly exceeds
$|\hat{c}_1|$, while for $r> 2.5$ it underestimates it.
At $r=2.7$, the $|c_{\rm rzif}|$ profile has a secondary minimum which
is absent from the corresponding $|\hat{c}_1|$ as well as from 
$|\theta_{\rm rzif}|$, $|\psi_{\rm rzif}|$, $|\hat{\theta}_1|$ and $|\hat{\psi}_1|$.
(The secondary minimum is, however, found in
$|\hat{c}_1|$ when $L$ is increased to 0.2.)
Figure \ref{fig:C_exactSCMfirstFourprofiles}(d) compares $||\hat{u}_1||=
\left[\int dz (|\hat{t}_1(z)|^2+|\hat{c}_1(z)|^2+|\hat{\psi}_1(z)|^2)\right]^{1/2}$
to its SCM approximation, including its higher order generalizations 
to be described in the next section.
We again see the overestimate by SCM of $|\hat{u}_1|$ for $r<2.5$ and its underestimate for
$r>2.5$.

\section{Higher-order methods}
\label{sec:higher-order}

We have seen that for the traveling waves of thermosolutal convection,
linearization about the full mean flow (RZIF) succeeds in matching 
the frequency of the nonlinear waves, while linearization about
a first-order approximation to the mean flow (SCM) does not.
It seems natural to consider whether higher-order approximations
to the mean flow can lead to a better match.

\subsection{Higher order SCM}

The SCM is a truncation of the Fourier decomposition of the exact system 
\eqref{eq:zerothFreq}-\eqref{eq:otherFreq}
including only components with $|n|\leq 1$.
A natural idea is to truncate at the next order, $|n|\leq 2$.
Meliga \cite{meliga2017harmonics} called this approximation
second-order SCM and implemented it for the flow over an open cavity,
using a multiple scale expansion method.
Truncating at this order, we obtain
\begin{subequations}
  \begin{align}
     0&=\mL\oU + \mN(\oU,\oU) + \mN({u}_1,{u}_{-1}) + \mN({u}_{-1},{u}_{1}) + \mN({u}_2,{u}_{-2}) + \mN({u}_{-2},{u}_{2}) 
     \label{eq:SCMmean2}\\
      i\omega {u}_1 &= \mL_{\oU}{u}_1 + \mN({u}_{2},{u}_{-1}) + \mN({u}_{-1},{u}_{2}) 
     \label{eq:SCMlsa21} \\
      2i\omega {u}_2 &= \mL_{\oU}{u}_{2}  + \mN({u}_1,{u}_1)      \label{eq:SCMlsa22}
  \end{align}
\label{eq:SYSscm2}
\end{subequations}
along with a phase condition (see section \ref{sec:algorithms}).
This system has as unknowns one real ($\oU$) and
two complex fields ($u_1$, $u_2$) and one unknown frequency ($\omega$).
In these equations,
$\oU$ does not signify the exact mean flow and the $u_n$'s do not
signify the exact Fourier components $\hu_n$ but 
approximations to them.
We call this truncated system SCM$_2$.
We can also extend \eqref{eq:SYSscm2} to include higher order terms,
forming third and higher order SCM approximations 
by truncating the exact representation \eqref{eq:zerothFreq}-\eqref{eq:otherFreq} at order $M$:\\
\begin{subequations}
  \begin{align}
  0 &= \mL\oU + \mN(\oU,\oU) + \sum_{1\leq |m|\leq M}\mN(u_m,u_{-m})
     \label{eq:SCMmeann}\\
      in\omega {u}_n &= \mL_{\oU}{u}_n + \sum_{1\leq |m|,|n-m|\leq M}\mN({u}_m,{u}_{n-m}) \qquad\qquad 1\leq n \leq M
     \label{eq:SCMlsan} 
  \end{align}
\label{eq:SYSscmn}
\end{subequations}

Higher order SCM does not fit into the category of the quasilinear or
semilinear models, since nonlinear interactions between
$\{u_1,u_2, \ldots\}$ that do not contribute to $\oU$ are included,
i.e. they are present in \eqref{eq:SCMlsa21}-\eqref{eq:SCMlsa22}
and in \eqref{eq:SCMlsan}. Instead, higher order
SCM, like harmonic balance, consists of a consistent truncation in
temporal modes at increasingly higher order. The optimal forcing
problem for a flat-plate boundary layer was solved at successively
higher orders of temporal frequency by \cite{rigas2021nonlinear}.

We solve system \eqref{eq:SYSscm2} or
\eqref{eq:SYSscmn} by a straightforward Newton's method (see section 
\ref{sec:algorithms}).
In these equations (and only here) we have been imprecise in
our notation; in theory, $\oU$, $u_n$ and $\omega$ should all carry labels
indicating that they are solutions of the 
$M^{\rm th}$ order system SCM$_{\rm M}$, but such labels would make these equations
unreadable.

\begin{figure}
    \includegraphics[width=4.3cm]{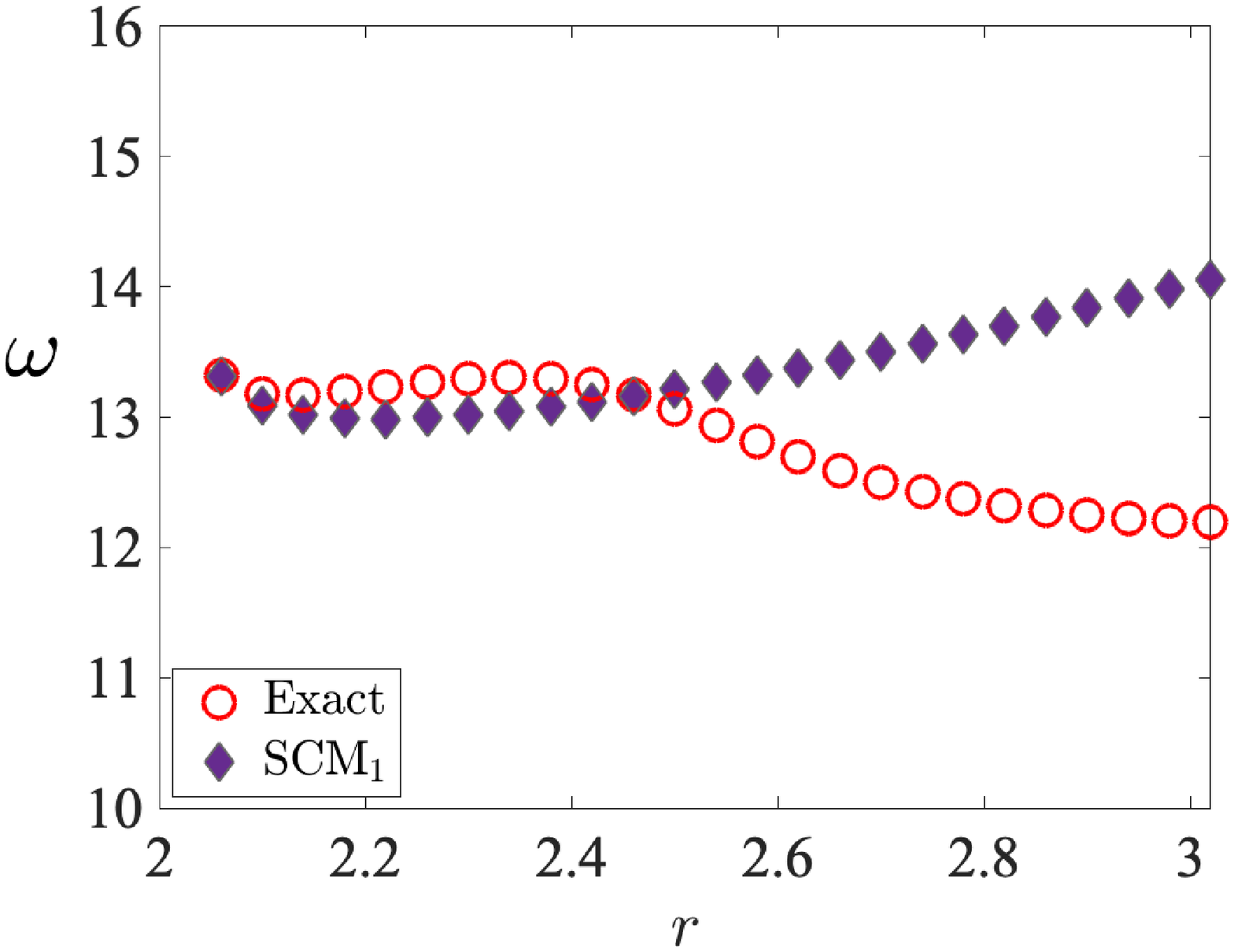}
    \includegraphics[width=4.3cm]{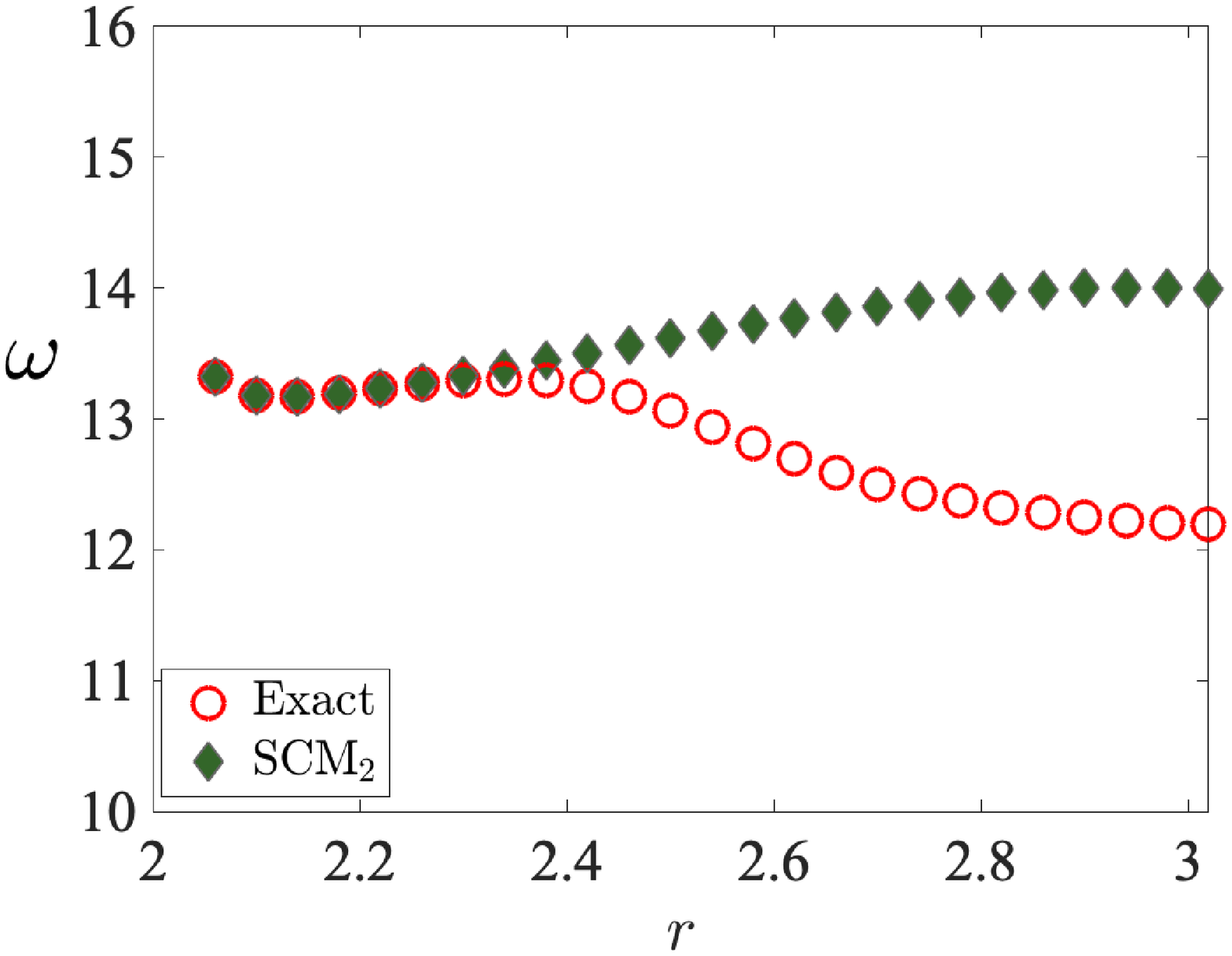}
      \includegraphics[width=4.3cm]{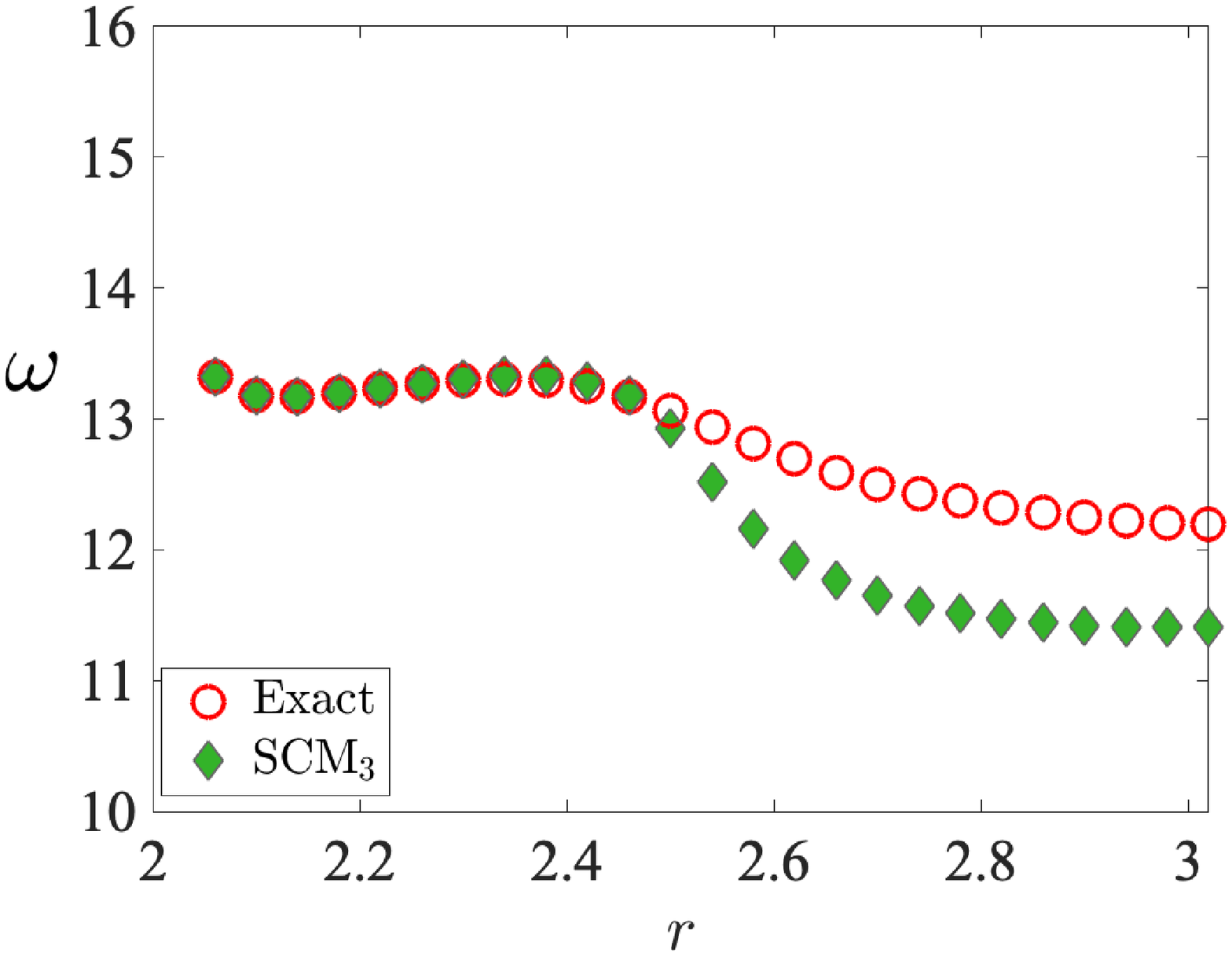}
      \includegraphics[width=4.3cm]{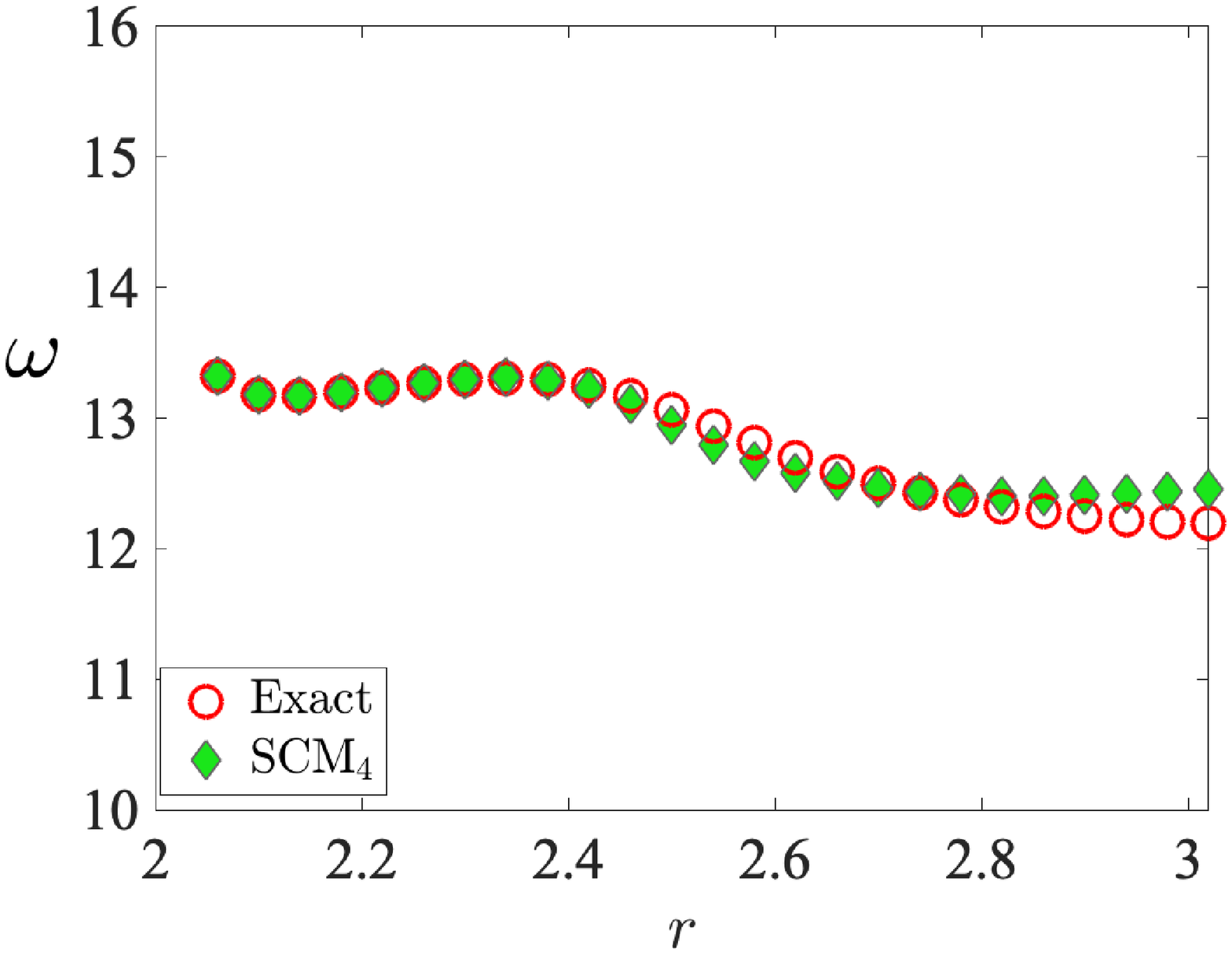}\\
\vspace*{-3cm}
{\it \hspace*{-4cm}(a) \hspace{3.9cm}(b)\hspace{3.9cm} (c)\hspace{3.9cm} (d)}
\vspace*{3cm}
      \caption{Frequency calculated by SCM methods of increasing order as a
        function of Rayleigh number. Exact frequencies are shown by
        open circles ($\color[rgb]{1.,0.,0.}\circ$), while those 
        predicted by the SCM are shown by diamonds: (a) first order
        (${\color[rgb]{0.5,0.,0.5}\blacklozenge}$), (b) second order
        (${\color[rgb]{0.2,0.4,0.1608}\blacklozenge}$), (c) third
        order (${\color[rgb]{0.2,0.702,0.1608}\blacklozenge}$), and
        (d) fourth order
        (${\color[rgb]{0.102,0.902,0.102}\blacklozenge}$).}
\label{fig:ScmOmg2345}
%
\includegraphics[width=4.3cm]{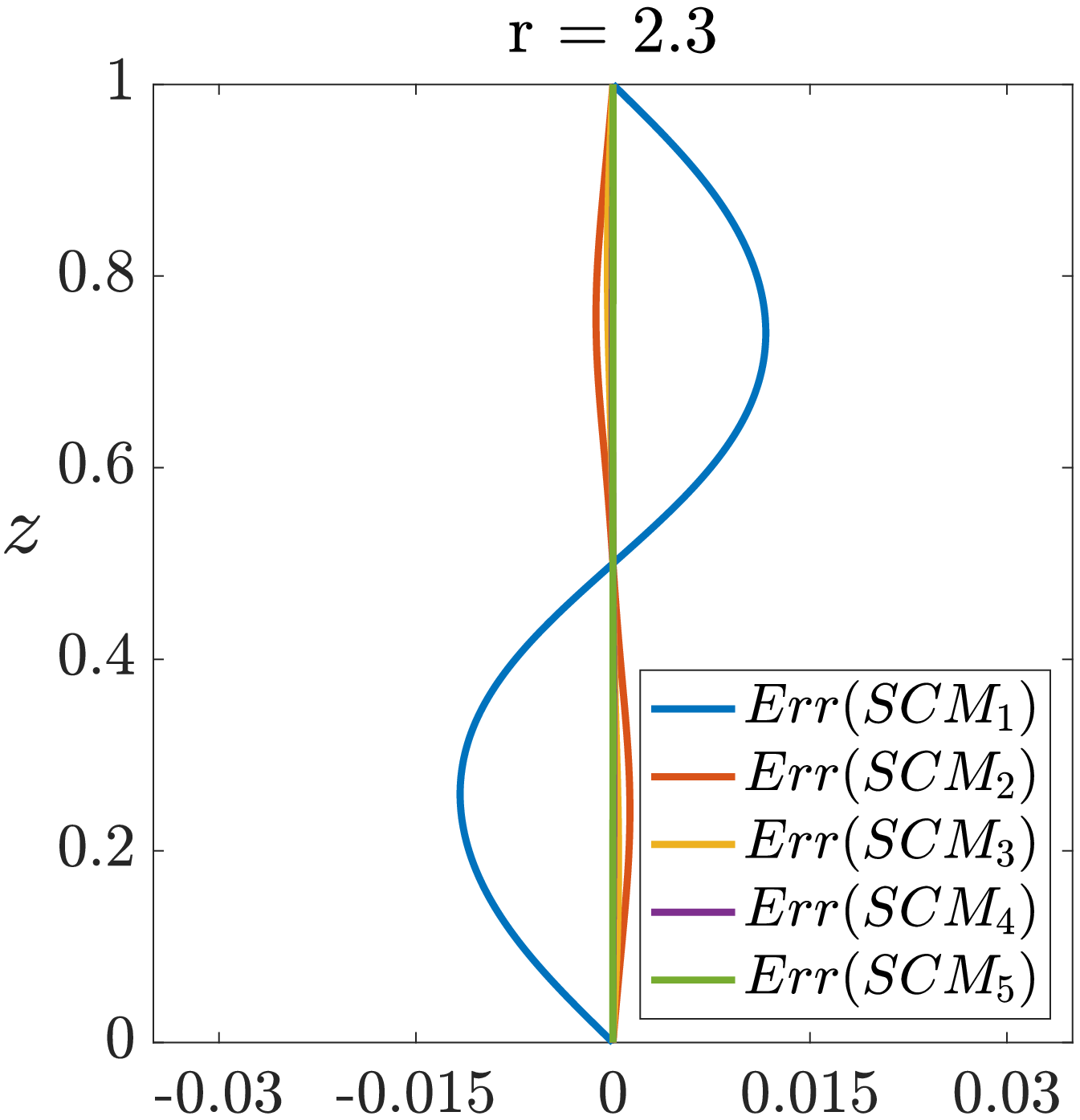}
      \includegraphics[width=4.3cm]{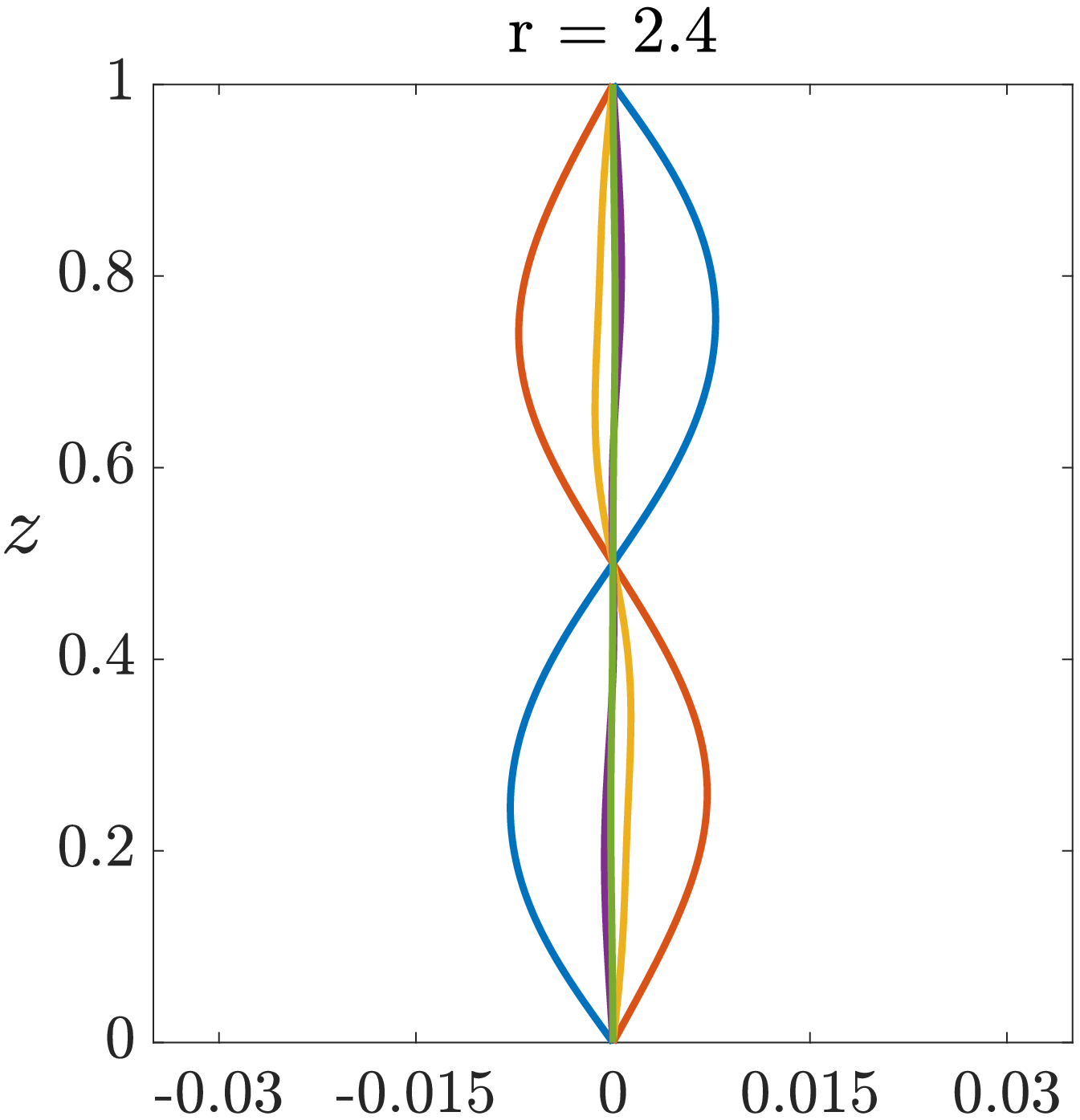}
      \includegraphics[width=4.3cm]{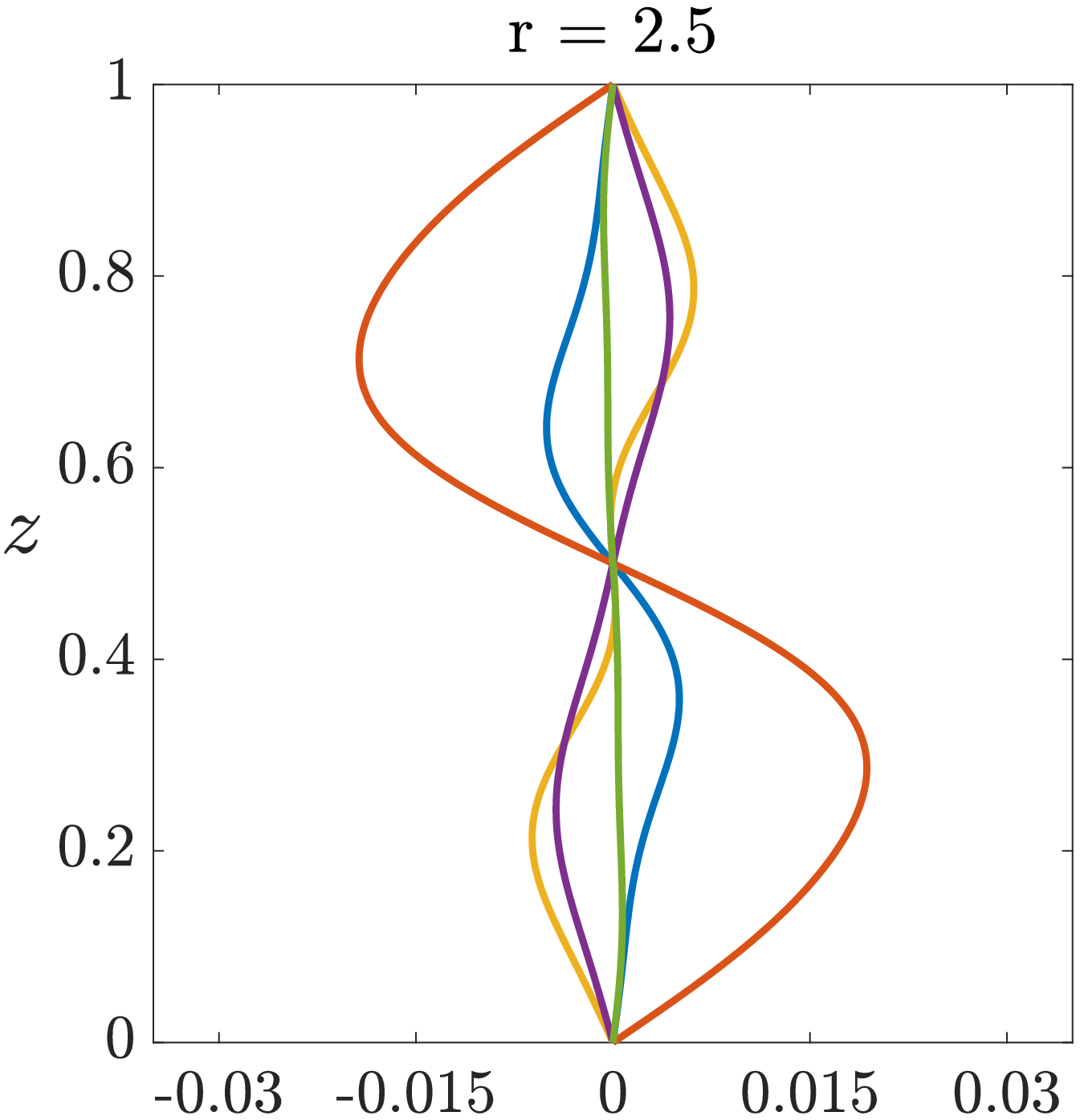}
      \includegraphics[width=4.3cm]{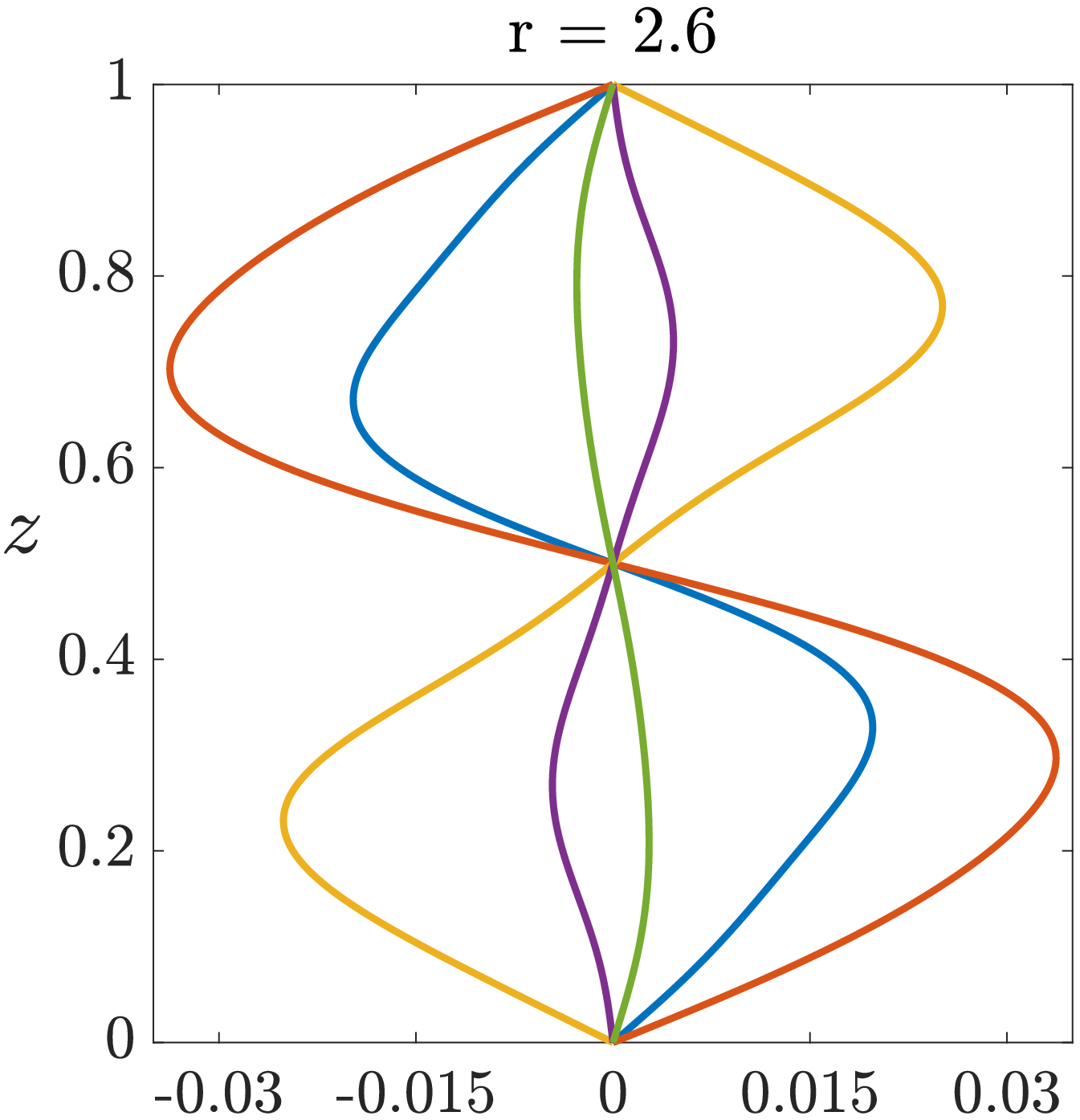}\\
\vspace*{-4cm}
{\it \hspace*{-4cm}(a) \hspace{3.9cm}(b)\hspace{3.9cm} (c)\hspace{3.9cm} (d)}
\vspace*{4cm}

      \caption{Mean concentration profiles calculated by various
        orders of the SCM method compared to exact mean concentration
        profile. (a) For $r=2.3$, (b) $r=2.4$, (c) $r=2.5$, and (d) $r=2.6$.}
        \label{fig:SCMprofiles}
%
\includegraphics[width=4.3cm]{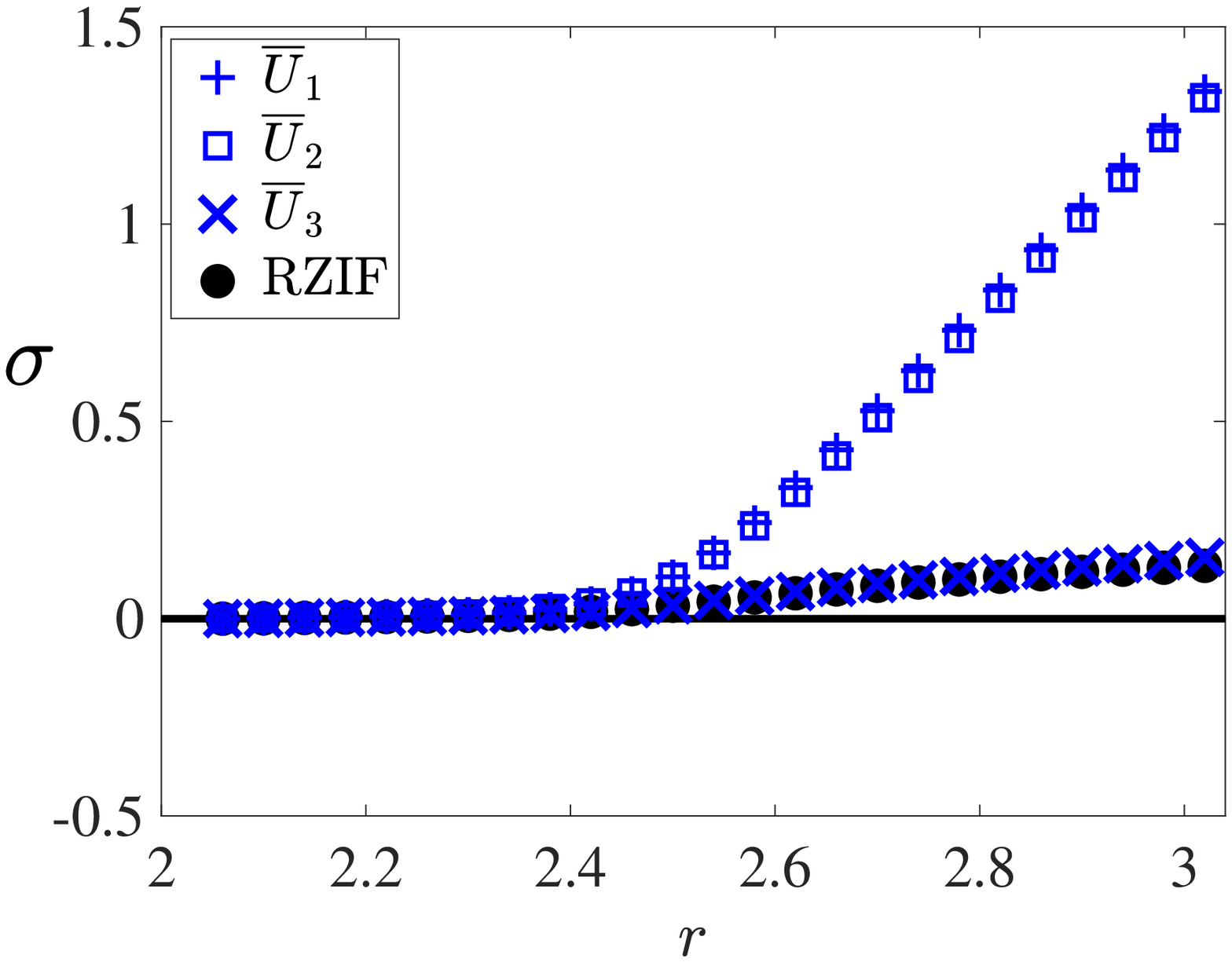}
\includegraphics[width=4.3cm]{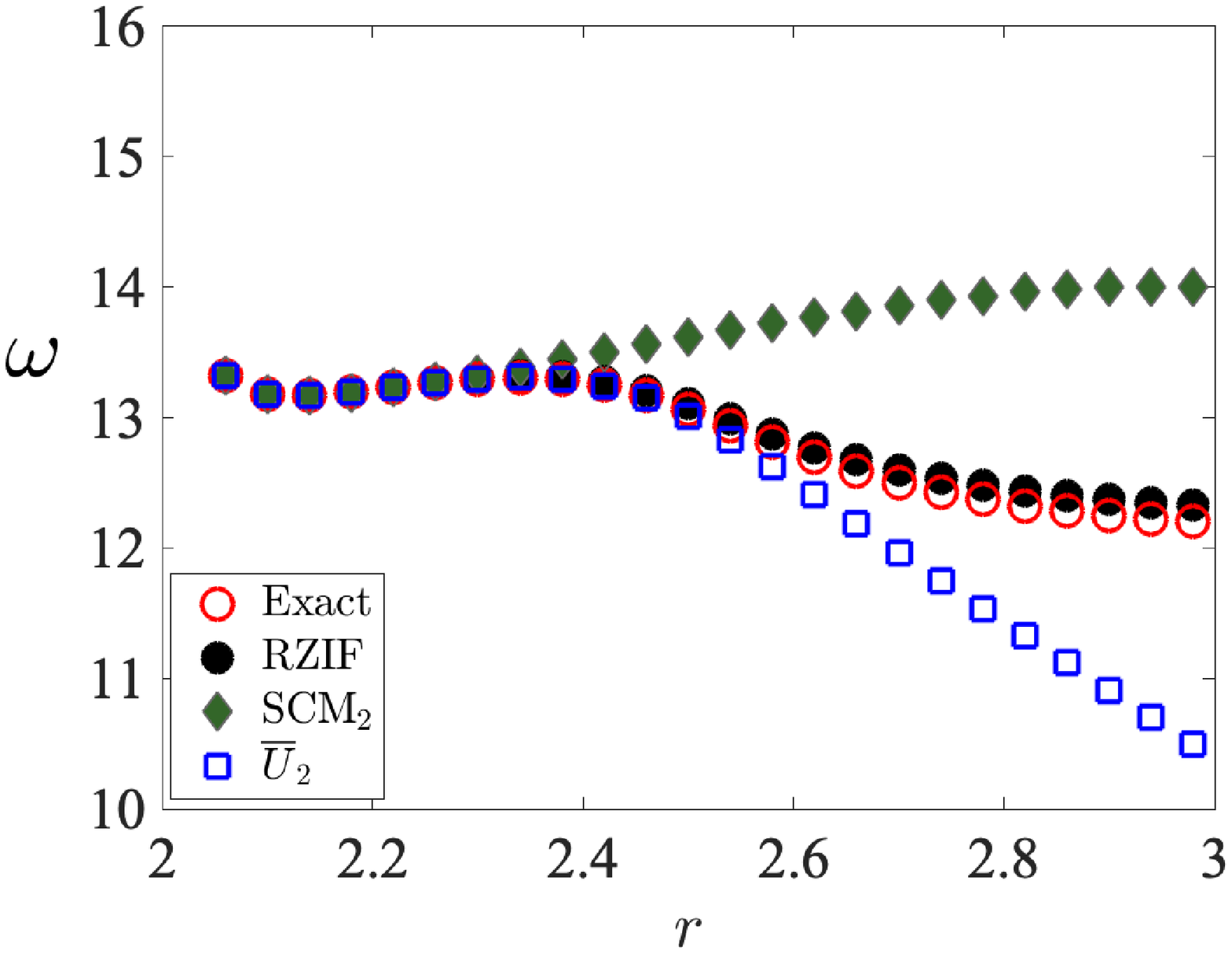}
  \includegraphics[width=4.3cm]{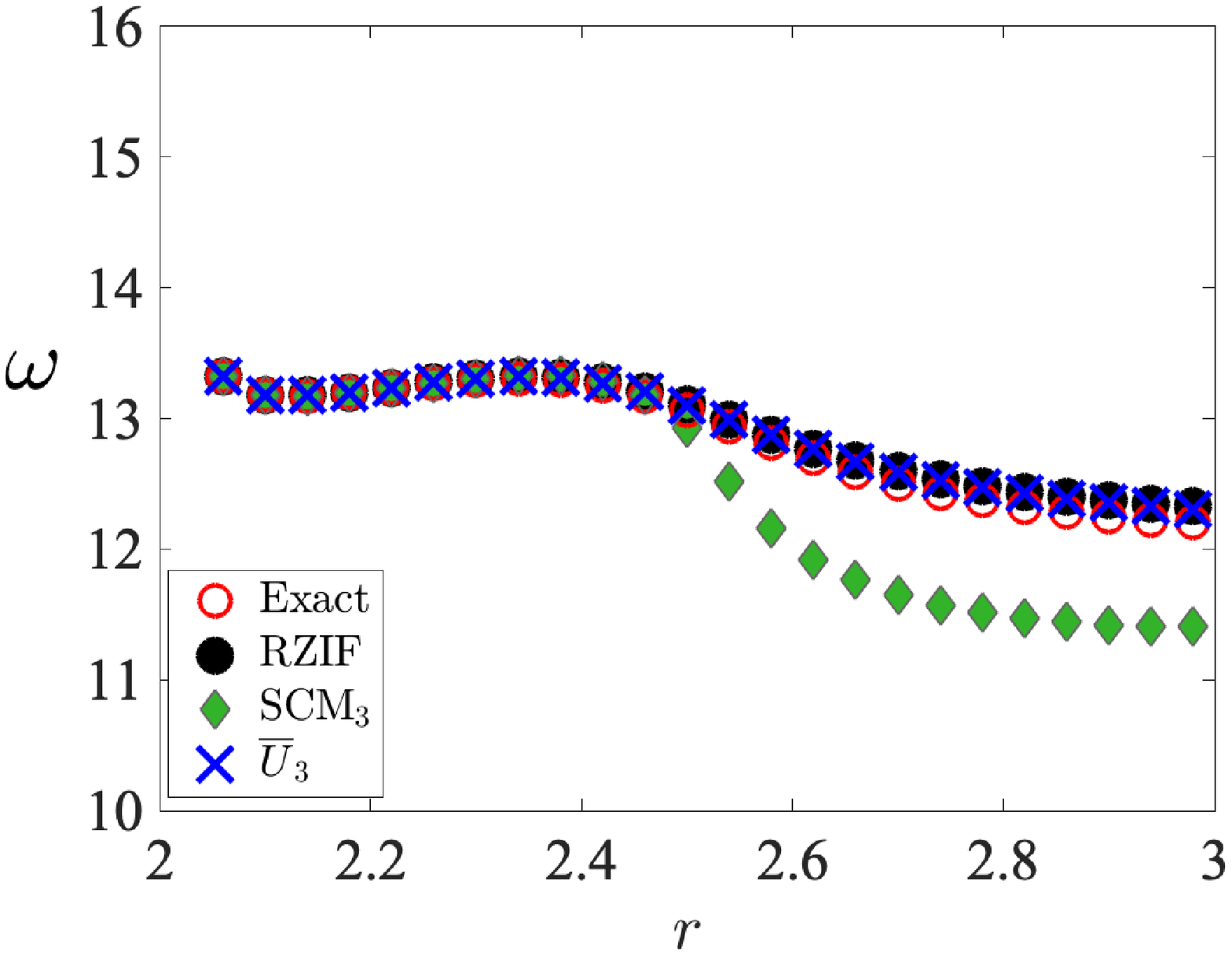}
  \includegraphics[width=4.3cm]{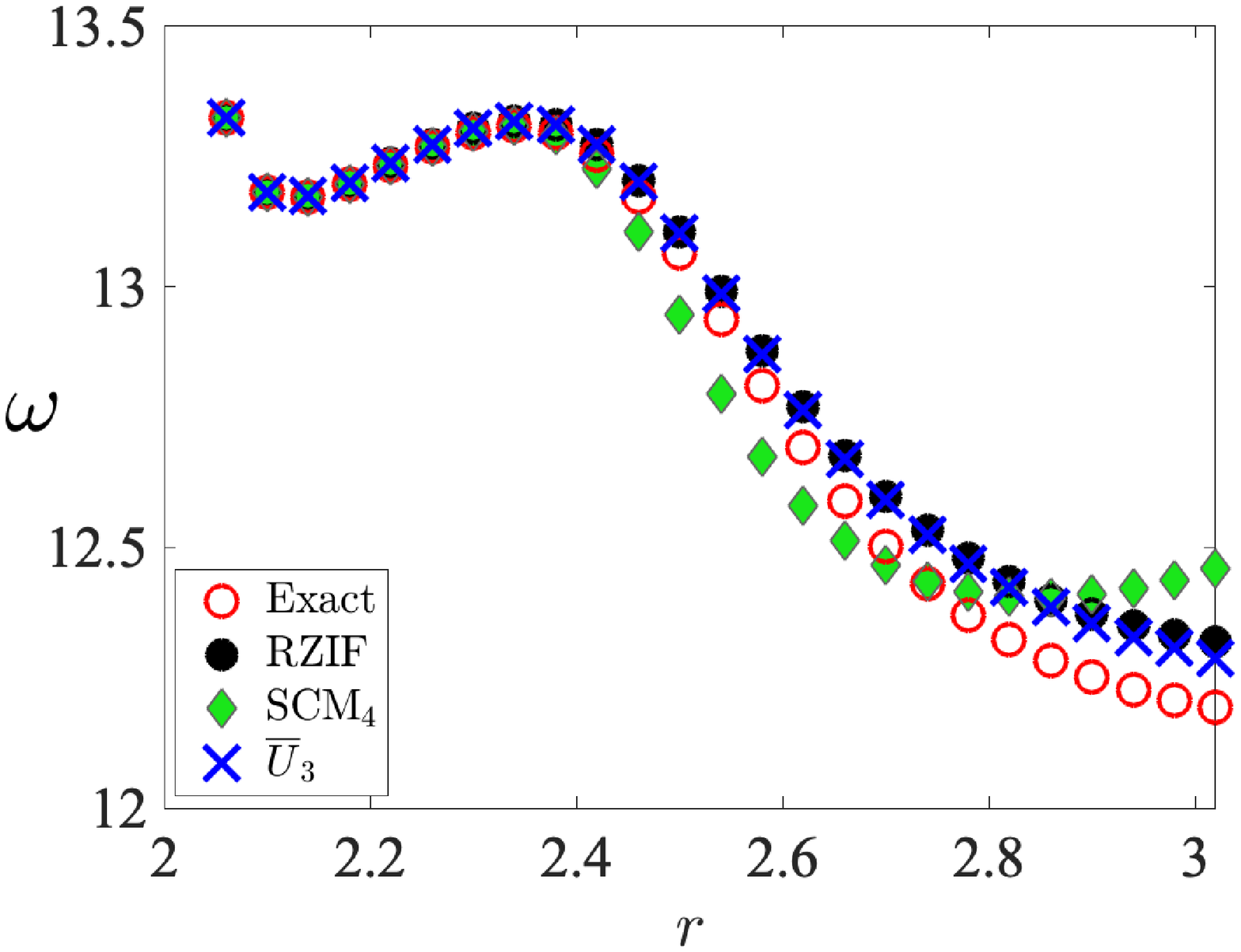}\\
\vspace*{-3cm}
{\it \hspace*{-4cm}(a) \hspace{3.9cm}(b)\hspace{3.9cm} (c)\hspace{3.9cm} (d)}
\vspace*{3cm}
    \caption{Comparison of the incomplete RZIF and SCM methods.
      (a) Growth rates calculated by linearizing about
      the full mean (RZIF) are represented by
      solid black circles ($\bullet$).
      Those obtained by linearizing
      about $\oU_1$, $\oU_2$ and $\oU_3$ are represented by blue plus signs
      ($\color[rgb]{0.,0.,1.} +$), by blue crosses 
      ($\color[rgb]{0.,0.,1.} \times$), and by hollow blue boxes, 
      ($\color[rgb]{0.,0.,1.} \Box$), respectively.
      The growth rates from $\oU_1$ and $\oU_2$ are almost indistinguishable, 
      as are the growth rates from $\oU_3$ and RZIF.
      (b,c,d) Frequencies.
      Exact frequencies are represented by open red circles
      ($\color[rgb]{1.,0.,0.}\circ$) while those obtained 
      by linearizing about the full mean field (RZIF) are represented by
      solid black circles ($\bullet$). 
      Those obtained by linearizing
      about $\oU_2$ and $\oU_3$ are represented by hollow blue boxes 
      ($\color[rgb]{0.,0.,1..} \Box$),
       and by blue crosses
      ($\color[rgb]{0.,0.,1.}\times$), respectively.
      Frequencies predicted by the SCM are shown by diamonds:
      (b) second order (${\color[rgb]{0.2,0.4,0.1608}\blacklozenge}$), 
      (c) third order (${\color[rgb]{0.2,0.702,0.1608}\blacklozenge}$), and
      (d) fourth order (${\color[rgb]{0.102,0.902,0.102}\blacklozenge}$).
      Linearization about $\oU_3$ achieves the same results as RZIF,
      so no further improvement is possible. SCM$_4$ is not as accurate.}
\label{fig:TwRecon}
%
\includegraphics[height=4.3cm]{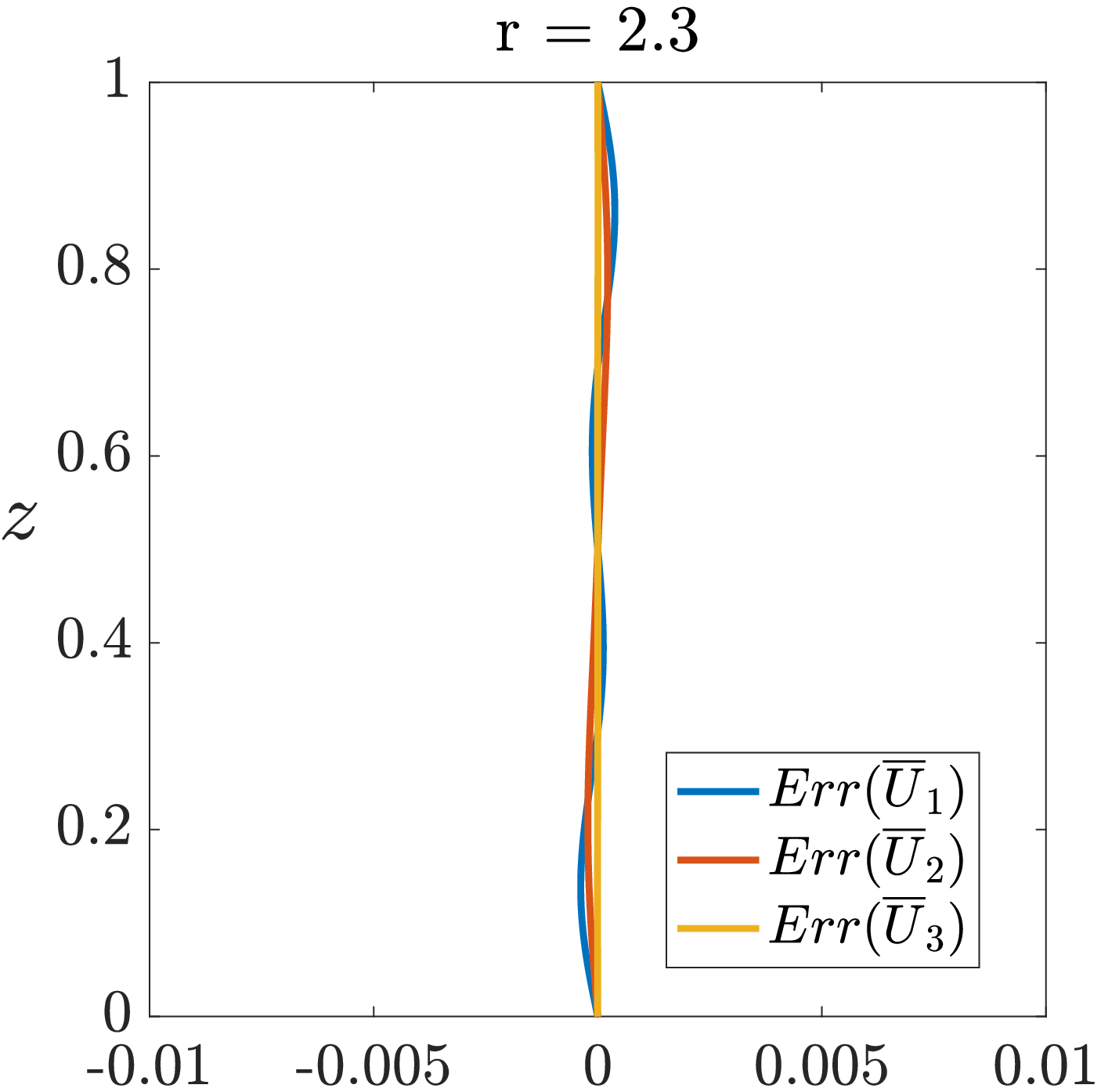}
\includegraphics[height=4.3cm]{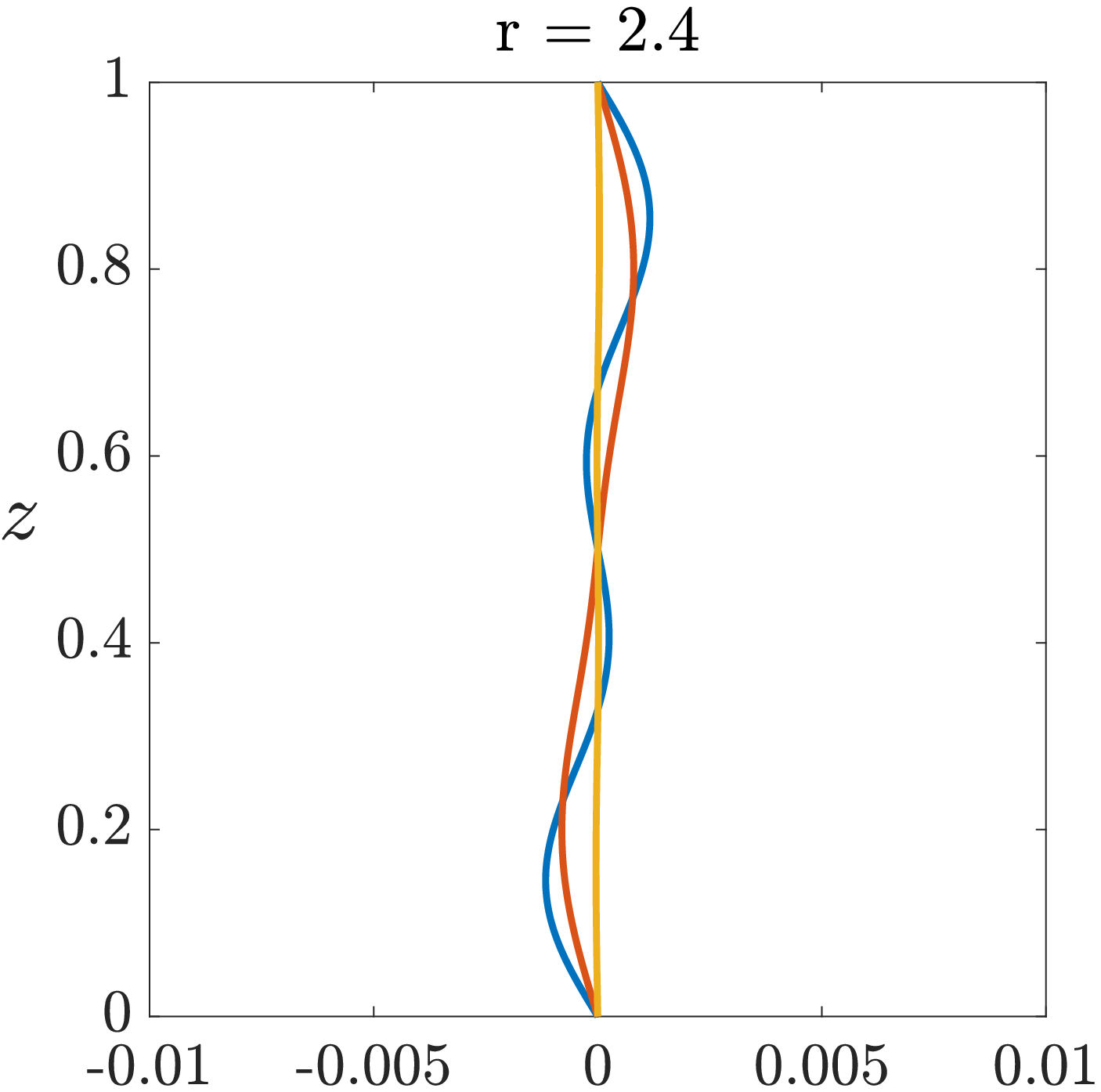}
\includegraphics[height=4.3cm]{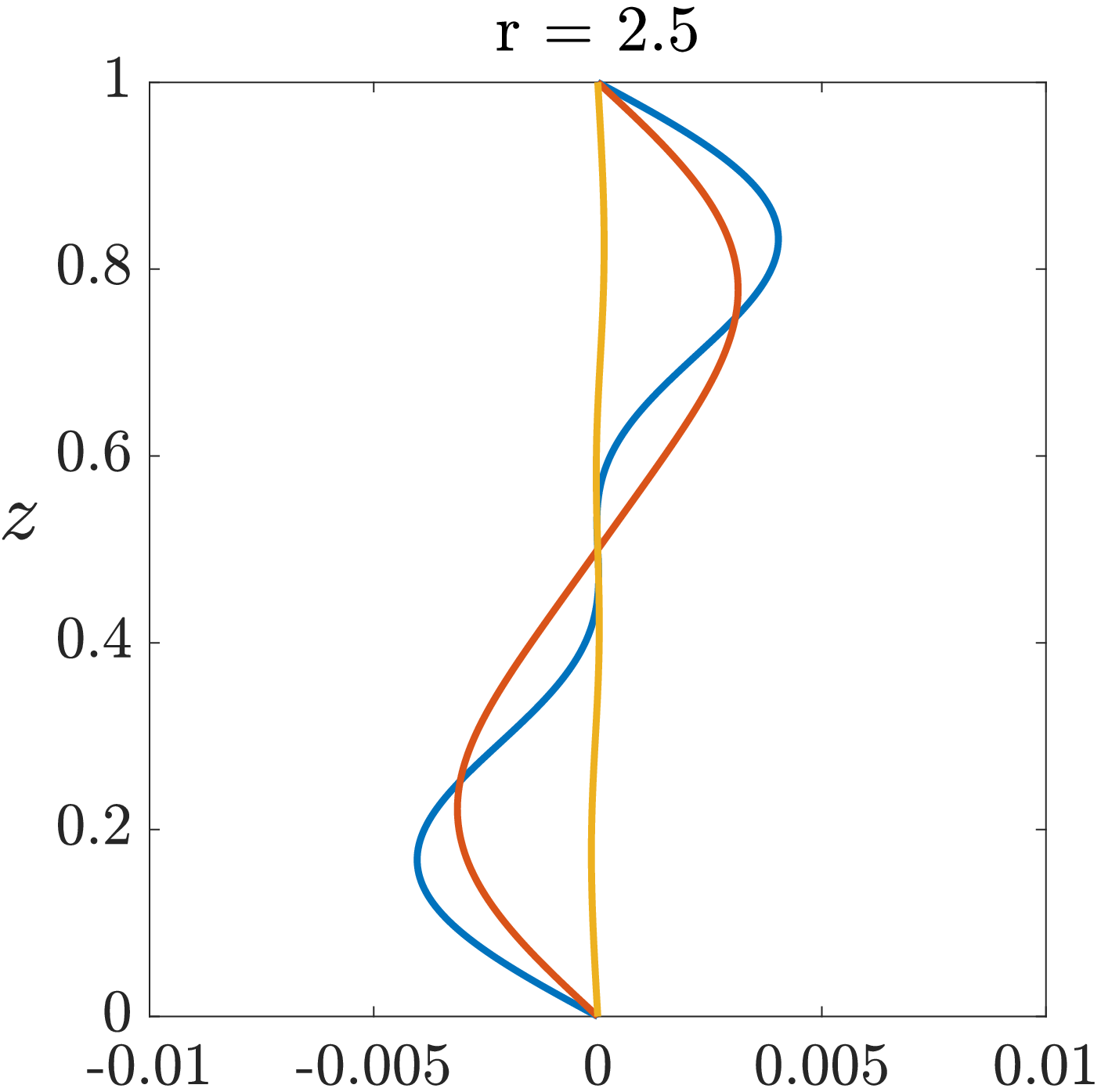}
\includegraphics[height=4.3cm]{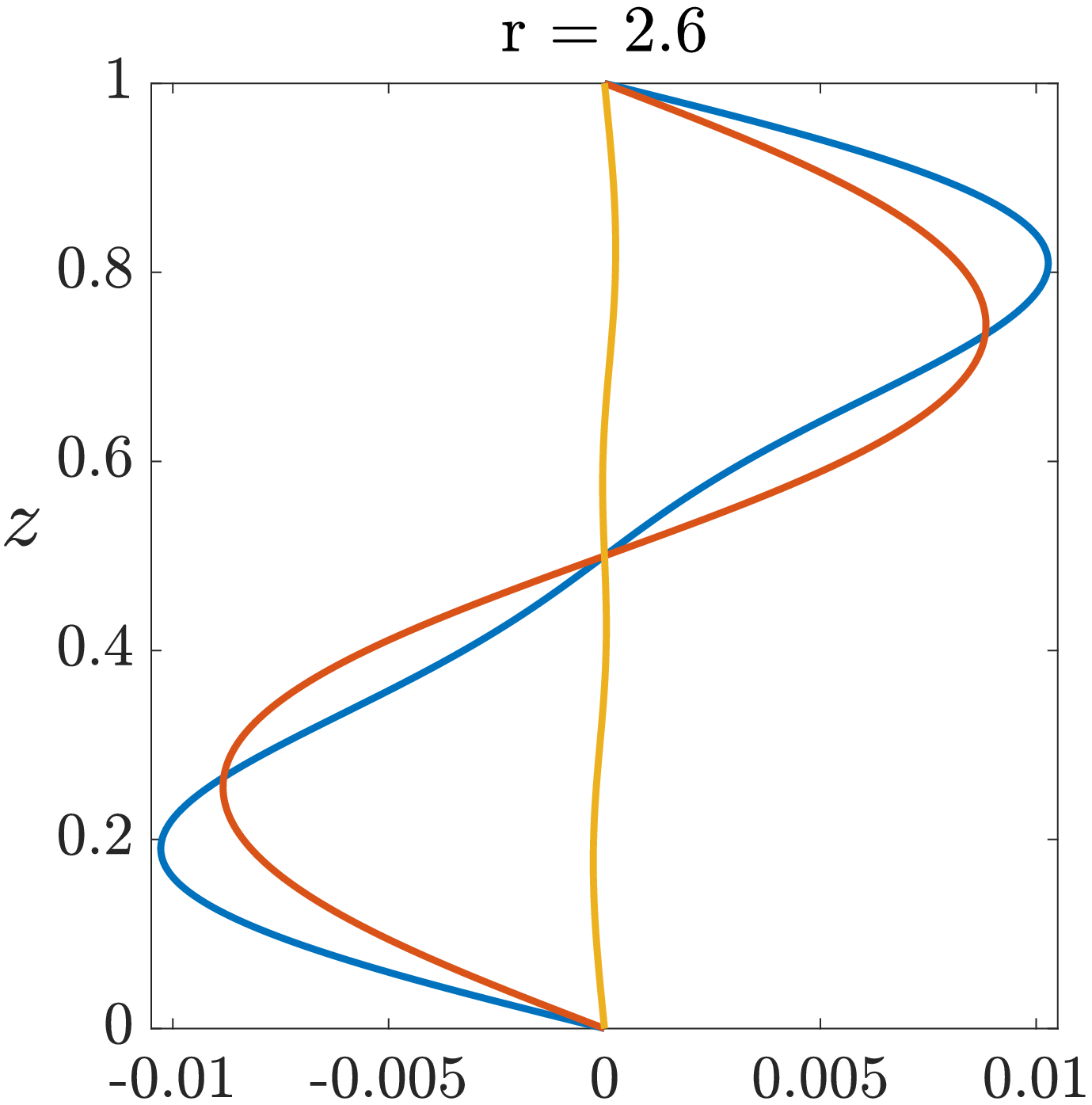}\\
\vspace*{-4cm}
{\it \hspace*{-4cm}(a) \hspace{3.9cm}(b)\hspace{3.9cm} (c)\hspace{3.9cm} (d)}
\vspace*{4cm}
      \caption{Mean concentration profiles
        calculated by various orders of the incomplete RZIF compared
        to exact mean concentration profile. (a) For $r=2.3$, (b)
        $r=2.4$, (c) $r=2.5$, and (d) $r=2.6$.  Note that the scale of
        the horizontal axis is one third that of figure
        \ref{fig:SCMprofiles}, indicating greater accuracy for the
        results of the incomplete RZIF method compared to the SCM.}
        \label{fig:buildupprofiles}
\end{figure}

Figure \ref{fig:ScmOmg2345} extends figure \ref{fig:ScmOmg1st}
by comparing the frequencies
computed by the higher order SCM systems with the exact frequencies.
Figure \ref{fig:ScmOmg2345}(b) shows that SCM$_2$
extends the range in which the frequency is well predicted from
$[2.05, 2.08]$ to $[2.05, 2.3]$, above which SCM$_2$
increasingly overestimates the frequency.
SCM$_3$ extends the matching range up to $r \approx 2.5$,
as shown in figure \ref{fig:ScmOmg2345}(c),
and underestimates the frequency above this range.
Figure \ref{fig:ScmOmg2345}(d) shows that SCM$_4$ considerably 
improves the frequency prediction throughout the $r$ range $[2.05, 3]$.
Since the SCM$_M$ equations converge to the exact equations with increasing
$M$, the corresponding frequencies must converge to the exact
frequencies.

Figure \ref{fig:SCMprofiles} extends figure \ref{fig:C_exactSCMmeanprofiles}
by presenting the error in the mean concentration profiles computed
by SCM$_M$ as $M$ is increased. 
Figure \ref{fig:SCMprofiles}(a) at $r=2.3$ shows the dramatic improvement
in the mean profile as $M$ is increased past 1,
as expected by comparing figures \ref{fig:ScmOmg2345}(a) and \ref{fig:ScmOmg2345}(b).
In contrast, figure \ref{fig:SCMprofiles}(b) at $r=2.4$ shows that
the deviation is as large for the SCM$_2$ profile
(and in the opposite direction) as it is for SCM$_1$. 
Figure \ref{fig:SCMprofiles}(c) at $r=2.5$ shows that, rather than improving
the profile, the SCM$_2$ approximation is even poorer than that of SCM$_1$.
The higher-order profiles converge to the correct profile, but non-monotonically.
This trend continues for $r=2.6$, shown in figure \ref{fig:SCMprofiles}(d).

\subsection{Incomplete RZIF}

The uneven performance of SCM has motivated us to perform another numerical
experiment, namely to build up the exact mean field by
truncating the contributions to it from the exact Fourier coefficients.
We denote these approximate mean fields 
by $\oU_1$, $\oU_2$, $\oU_3, \ldots$ and linearize about them:
\begin{subequations}
\begin{align}
  0 &= \mL\oU_M + \mN(\oU_M,\oU_M) + \sum_{1\leq |m|\leq M}\mN(\hu_m,\hu_{-m})
       \label{eq:constructmean}\\
  (\sigma_M+i\omega_M)u_M &= \mL_{\oU_M}  u_M
     \label{eq:constructu}
\end{align}
\end{subequations}
where the $\hu_m$ contributing to the mean $\oU_M$ in \eqref{eq:constructmean}
are the exact Fourier components of the nonlinear limit cycle defined
in \eqref{eq:DecompFourier}.  In our case,
$\mN(\oU_M,\oU_M)=\mN(\oU,\oU)=0$, so \eqref{eq:constructmean} can be
solved via
\begin{align}
  \oU_M &=-\mL^{-1} \sum_{1\leq |m|\leq M}\mN(\hu_m,\hu_{-m})
\end{align}
We will call this the incomplete RZIF approximation.

It is useful to compare this system with the higher order SCM$_M$
system \eqref{eq:SCMmeann}-\eqref{eq:SCMlsan} and with the exact system
\eqref{eq:zerothFreq}-\eqref{eq:otherFreq}.
Although equation \eqref{eq:constructmean} resembles
\eqref{eq:SCMmeann}, we emphasize that the exact Fourier components
${\hu}_n$ of the nonlinear limit cycle $\ULC$ are used in
\eqref{eq:constructmean}, as they are in the corresponding exact
equation \eqref{eq:zerothFreq}.  In contrast, the SCM equation
\eqref{eq:SCMmeann} uses approximate Fourier components defined
self-consistently by the coupled truncated system
\eqref{eq:SCMmeann}-\eqref{eq:SCMlsan}.
On the other hand, \eqref{eq:constructu} omits all of the terms
$\mN_1,\mN_2,\ldots$, as in RZIF, whereas \eqref{eq:SCMlsan} includes
increasingly accurate versions of these terms.  Thus, the incomplete
RZIF approximation \eqref{eq:constructmean}-\eqref{eq:constructu} is a
gradual approach to RZIF rather than to the full exact equations
\eqref{eq:zerothFreq}-\eqref{eq:otherFreq}.  The incomplete RZIF
approximations of various orders are less accurate than the original
RZIF method of sections \ref{sec:framework}-\ref{sec:applic}, in
contrast to the SCM$_M$ methods of various orders, which are more
accurate than the original SCM$_1$ method.

Figure \ref{fig:TwRecon} shows the eigenvalues resulting from the
incomplete RZIF approximation.  First, figure \ref{fig:TwRecon}(a)
shows the real parts $\sigma_M$ as a function of $r$. For $r < 2.5$,
$\sigma_M \approx 0$, but for $r > 2.5$ and for $M=1$ and $M=2$, the
values of $\sigma_M$ are quite far from zero. Note that
$\sigma_2 \approx \sigma_1$, implying that adding the contribution
from $\mN(\hu_2,\hu_2)$ does not improve the estimated mean flow
$\oU_2$.  This is also true for the imaginary parts:
$\omega_2 \approx \omega_1$.  In figure \ref{fig:TwRecon}(b), we
compare $\omega_2$ to the exact frequency $\oLC$, the RZIF frequency
$\oR$, and the frequency from SCM$_2$.  The estimates $\omega_1$ (not
shown in the figure) and
$\omega_2$ are fairly accurate for $r\leq 2.6$, whereas the frequency
from SCM$_2$ is accurate only for $r\leq 2.4$.  For $M=3$, Figure \ref{fig:TwRecon}{c}
shows that the
frequencies $\omega_3$ are almost indistinguishable from $\oR$ and
$\oLC$, while those from SCM$_3$ still deviate for $r\geq 2.5$.  Note
that $\omega_M$ cannot exceed the accuracy of $\oR$, since the terms
$\mN_1$, $\mN_2$, etc. continue to be neglected.  This is emphasized
in the enlargement of panel (d), where $\omega_3$ is very close to
$\oR$ while remaining apart from $\oLC$.  For $r>2.4$, the frequency
from SCM$_4$ follows a different trend.

Figure \ref{fig:buildupprofiles} shows the error in the mean concentration
profiles resulting from successively truncating the Fourier series, as
in \eqref{eq:constructmean}. These errors are considerably smaller than 
the corresponding errors from the SCM analysis; 
the scale of figure \ref{fig:buildupprofiles} is a third
of that of figure \ref{fig:SCMprofiles}. We see that
going from $\oU_1$ to $\oU_2$ does not substantially decrease the error
in the incomplete RZIF approximation,
while $\oU_3$ achieves the accuracy of RZIF, as was seen in figure
\ref{fig:TwRecon} for the eigenvalues.

The incomplete RZIF approximation removes the effect of approximating the Fourier components,
leaving only the effect of truncating the Fourier sum.
The less satisfactory performance of SCM compared to the incomplete RZIF
method of the same order can thus be attributed to the inaccuracy
in SCM's estimates of $\hu_1, \hu_2, \ldots$, leading to inaccuracy in
the estimated mean flow.
Including higher-order modes produced by self-consistent truncations
proves less successful than including their exact versions 
at the same order.

We mention that neither of the families of methods -- higher-order SCM
nor incomplete RZIF -- fall precisely into the category of QL or GQL methods.
We recall that QL or GQL methods divide the modes into two types, the
mean (or low frequency) modes and the other (or high frequency) modes.
One set of equations involves only the projections onto low modes of
low-low or high-high quadratic terms. 
The other set involves only mixed low-high quadratic terms, so that 
the high frequency modes obey equations which are linear in the high frequency terms.
In contrast, the RZIF methods use externally calculated (exact) fields
while the SCM methods include all interactions between the retained modes.

\section{Algorithms}
\label{sec:algorithms}

\subsection{Thermosolutal convection}
\label{sec:alg_conv}
We first describe the methods particular to thermosolutal convection.
The spatial discretization consists of a Fourier series in the
periodic direction $x$ and a sine series in the vertical direction $z$
(allowed for the streamfunction because of the free-slip boundaries).
Differentiation is carried out in Fourier-sine space and multiplication
in the grid space. For our parameter range and boundary conditions, 
very little resolution is needed; the $(x,z)$ rectangle is represented
by a $16 \times 8$ grid. 
By defining 
\begin{align}
U \equiv(\Temp, C, \Psi)^T
\end{align}
we rewrite \eqref{eq:goveq} in the compact notation used previously
\begin{align}
\pd_tU = \mL U + \mN(U,U)
\label{eq:compact}\end{align}
We carry out time evolution by a mixed scheme,
in which diffusive terms $\mL$ are evolved via the implicit Euler method
and the remaining terms by the explicit Euler method.
\begin{align}
  U(t+\dt) &= (I -\dt \mL)^{-1} \left[U(t)+\dt \mN(U(t),U(t))\right]
 \label{eq:BEFE}
\end{align}

When the limit cycle $\ULC$ is a traveling wave, 
it is a stationary state in a moving reference frame governed by
\begin{align}
V \pd_x \ULC  &=  \mL \ULC + \mN(\ULC,\ULC)
\label{eq:LCsteady}\end{align}
where $V =\lambda/\TLC=\oLC/k$ is the wavespeed, with $\lambda$ the
wavelength, $\TLC$ the period, $k$ the wavenumber, and $\oLC$ the
angular frequency.
The term $V\pd_x U$ can be moved to the
right-hand-side and integrated explicitly along with $\mN$.  The
traveling waves are computed via Newton's method by transforming 
\eqref{eq:BEFE} as described in \cite{turton2015}, with
time stepping providing initial estimates for fields and wavespeeds. 
To compensate for the additional variable of the wavespeed $V$, 
a phase condition such as
\begin{align}
  \pd_x \widetilde{U}_{\rm lc}(x=0)=0
\label{eq:phase}
\end{align}
is imposed, where $\widetilde{U}$ is taken to be one of $\Theta, C, \Psi$ at a fixed value of $z$. The traveling wave solution is continued
from one value of $r$ to the next in order to cover the range
$[2.06,3]$.

When the limit cycle is not a traveling wave, as is the case for the
cylinder wake or the standing waves of thermosolutal convection,
it must be calculated via time integration. Another
possibility is to use Newton's method with shooting to redefine the
limit cycle as a fixed point problem in a much higher dimensional
space.

\subsection{RZIF and SCM systems}
We now discuss algorithmic aspects specific to the 
RZIF and SCM equations. 
For RZIF, the limit cycle solution is averaged over time
(or equivalently, for a traveling
wave, over the $x$ direction) to produce $\oU$.
%
The Jacobian about $\oU$ is computed and
diagonalized to produce its leading eigenvalue $\sR+i\oR$. For this
small problem, matrix operations such as diagonalization and inversion
for Newton's method can be carried out directly, but for larger
problems, matrix-free iterative methods such as BiCGSTB, GMRES, or IDR
and the Arnoldi or power methods can be used.

We now turn to the SCM:
\begin{subequations}
\begin{align}
0 &= \mL \US +\mN(\US,\US) + \mN(\uS,\uS^*) \label{eq:SCM1}\\
i\oS\uS &= \mL_{\US} \uS \label{eq:SCM2}
\end{align}
together with a phase condition.
The unknowns are the real field $\US$, the complex field $\uS$,
and the scalar $\oS$.
We solve the coupled system \eqref{eq:SCM1}-\eqref{eq:SCM2}
via a straightforward Newton's method.
We start near the threshold $r=r_{\rm Hopf}$, where $\US = \Ub$
(which is zero in the thermosolutal case) and $\uS=\ub, \oS=\oL$.
For higher $r$ values,
the initial estimate used is the solution at the previous value of $r$.

Manti\v{c}-Lugo {\it et al.}~\cite{mantivc2014self,mantivc2015self}
solve the SCM equations by an 
iterative algorithm that decouples the two equations.
Equation \eqref{eq:SCM1} is treated as a nonlinear equation for $\US$
with $\mN(\uS,\uS)$ as an inhomogeneous forcing term,
while \eqref{eq:SCM2} is treated as an eigenproblem
with fixed $\US$ defining the linear operator.
As it stands, \eqref{eq:SCM2} is not an eigenproblem, since
$\mL_{\US}$ is expected to have complex eigenvalues rather than pure imaginary ones.
(The closely related operator $\mL_{\Ub}$ has an imaginary eigenpair
only exactly at the Hopf bifurcation.)
In addition \eqref{eq:SCM2} does not fix a normalization for $\uS$,
which is required for $\mN(\uS,\uS)$ when it is used as an input for
\eqref{eq:SCM1}.
Such considerations lead these authors to 
specify a norm $A$ for $\uS$ (or, equivalently, to multiply a normalized
$\uS$ by $A$). Equation \eqref{eq:SCM2} is replaced by 
\begin{align}
  (\sS+i\oS)\uS &= \mL_{\US} \uS \label{eq:SCM3}\\
  ||\uS||&=A \label{eq:SCM4}
\end{align}
\label{eq:SCM34}
\end{subequations}
where $\sS$ and \eqref{eq:SCM4} are 
an additional unknown and equation relative to \eqref{eq:SCM2},
while $A$ is an input value.

Determining $\US$, $\uS$, and $\oS$
for a single value of $r$ requires looping over values of $A$
as follows.
$A$ is initially set to zero, since then \eqref{eq:SCM1} and
\eqref{eq:SCM3} are the equations governing the base flow and leading
eigenpair from classical linear stability analysis; their solution is
$\Ub$, $\sL+i\oL$, $\ub$.
In order to solve the equations for a new
$A>0$, $\uS$ is given norm $A$ and substituted into
\eqref{eq:SCM1} to generate a new $\US$, which is in turn substituted
into \eqref{eq:SCM3}-\eqref{eq:SCM4}, leading to a new $\uS$
that is substituted into \eqref{eq:SCM1}. The process is continued until
$\US$, $\sS+i\oS$ and $\uS$ cease to change.  $A$ is then increased
and the procedure repeated, using as initial estimates the solutions
for the previous $A$.  The calculation is halted and the
solution accepted when a value of $A$ is reached for which $\sS=0$.
Thus, the Real Zero portion of the RZIF hypothesis is
built into the method. 

However, even if \eqref{eq:SCM1} and \eqref{eq:SCM3}-\eqref{eq:SCM4} can be
individually satisfied, there is no guarantee of convergence of the
coupled system for a given $A$.  Nor is it guaranteed that there will be a value of
$A$ such that $\sigma(A)=0$.  When Manti\v{c}-Lugo {\it et
  al.}~\cite{mantivc2014self,mantivc2015self} used the decoupled
algorithm to compute the SCM approximation for the cylinder wake, they
reported convergence problems, in response to which they introduced a
relaxation factor and a different normalization of $\mN(\uS,\uS)$ to
improve convergence; more details about the algorithm can be found in
\cite{mantivc2015self,mantivc2015phd}.  With these modifications, 
they were then able to accurately reproduce the frequency
of the cylinder wake for Reynolds numbers up to $Re=120$.

Meliga \cite{meliga2017harmonics} implemented the second order SCM$_2$
given by \eqref{eq:SYSscm2}
by generalizing the approach in \cite{mantivc2014self,mantivc2015self},
writing a series of nested sub-problems for 
$\oU$, $u_1$, $u_2$ and two auxiliary complex fields,
each solved via Newton's method and the Arnoldi method.
As in \cite{mantivc2014self,mantivc2015self}, an amplitude 
$A$ was imposed and the solution was considered to be
reached when a growth rate reached zero.

In our case of traveling waves in thermosolutal convection, we were
able to use the decoupled algorithm \eqref{eq:SCM1} and
\eqref{eq:SCM3}-\eqref{eq:SCM4} for $r$ only $2\%$ above $r_{\rm Hopf}$; above
this value, the decoupled algorithm does not converge.  In contrast,
the full Newton method performed robustly for \eqref{eq:SCM1}-\eqref{eq:SCM2},
as well as for the higher order SCM systems
\eqref{eq:SCMmeann}-\eqref{eq:SCMlsan}.
We note that Fani {\it et~al.} \cite{fani2018computation}
also applied a full Newton method to solve the SCM for the acoustic
generation of the B\'enard–-von K\'arm\'an vortex street, using
MUMPS to solve the large sparse linear system required by Newton's method.
We have presented the coupled algorithm for several reasons:\\
(i) We wished to make contact with the literature. \\
(ii) Our thermosolutal problem is quite small.
Although our method is considerably simpler, it is possible
that a decoupled method such as that in
\cite{mantivc2014self,mantivc2015self} or \cite{meliga2017harmonics} might 
be needed for a larger problem.\\
(iii) The decoupled algorithm has the advantage of describing the amplitude
saturation process, mimicking the evolution of $A$ in time, discussed
in \cite{maurel1995mean,zielinska1997strongly}. The unstable base
field solution extracts energy from the perturbations, which grow
until they saturate. SCM computes
the mean field, the nonlinear frequency and the nonlinear mode along with
its amplitude $A$ without time integration. 

\vspace*{-0.5cm}

\section{Conclusion}

Nonlinear equations can be interpreted as
governing the coupled evolution of modes, canonically Fourier modes. 
Various truncations have been proposed
in order to either speed up computations or to gain
a greater understanding of the behavior of their solutions. 
A basic task, which may be considered to be a benchmark
of such truncations, is to match the frequency of a limit cycle.

RZIF consists of computing the temporal
mean, linearizing the evolution operator about it,
and then calculating its leading eigenvalue \cite{barkley2006linear}.
This approximation 
has been shown to be resoundingly successful in
the archetypal case of the wake of the circular cylinder
\cite{barkley2006linear}, the traveling waves of thermosolutal
convection \cite{turton2015}, the ribbons and spirals of
counter-rotating Taylor-Couette flow \cite{bengana2019spirals}, and
the shear-driven flow over a square cavity \cite{bengana2019bifurcation}.
Although RZIF has thus far been applied only to limit cycles produced
by supercritical Hopf bifurcations, it is plausible that
it might also apply when the bifurcations are subcritical,
since the mean upon which it relies is obtained from the
nonlinear limit cycle, independently of its distance from the base flow.
The search for a general reason for this success
is constrained by the existence of a clear counterexample:
the standing waves of thermosolutal convection that bifurcate
at the same parameter value as the traveling waves
\cite{turton2015}. Based on this counterexample,
Turton et al.~\cite{turton2015} proposed that
the dominance of the primary Fourier mode could
serve as a criterion for success of RZIF, which pushes the question 
further upstream to when and why the primary Fourier mode dominates.

RZIF confers theoretical insight but no practical advantages, since
the temporal mean is calculated from a full simulation of the limit
cycle.  For this reason, Manti{\v{c}}-Lugo et
al.~\cite{mantivc2014self,mantivc2015self} proposed to close the
equations by limiting them to the mean flow and the primary temporal
Fourier mode and showed that this SCM method succeeded as dramatically
as RZIF on the archetypal cylinder wake.
%
However, figure \ref{fig:ScmOmg1st} shows that the traveling waves of
thermosolutal convection that satisfy the RZIF property so well cannot
be approximated by the SCM.  Although the interaction between
higher-order modes may be omitted from the higher-order equations
\eqref{eq:otherFreq}, their contribution to the mean flow remains
important: they cannot be removed from equation \eqref{eq:zerothFreq}
governing the mean flow.  In addition, in our example, a good
approximation of the mean flow requires that the higher-order
modes contributing to it be accurately represented, as demonstrated by
figures \ref{fig:ScmOmg2345} and \ref{fig:TwRecon}.  From this
example, it would seem to be interactions, rather than modes, that can
be omitted.
%
However, other examples, e.g.~\cite{mantivc2016saturation}, argue
in the opposite direction.


Despite verifying RZIF, the traveling waves of our thermosolutal
convection problem verify SCM in a very narrow interval around one
parameter value and not elsewhere.
The thermosolutal
standing waves that provide a counter example to RZIF are generated at
precisely the same bifurcation and with the same parameter values as
the traveling waves.  These counter examples provide a warning that
truncations must be carefully controlled and understood, and that
doing so may prove unexpectedly difficult.

From the example of the cylinder wake
\cite{mantivc2014self,mantivc2015self} and its compressible version
\cite{fani2018computation} as well as the shear-driven cavity
\cite{meliga2017harmonics} it is clear that the SCM works remarkably
well even for fairly complex hydrodynamic problems, while yielding a
major reduction in calculation costs.  The challenge is to determine
which configurations are amenable to SCM and why.

\bibliography{BiBthese.bib}
\end{document}